\documentclass[fleqn,usenatbib,useAMS]{mnras}

\usepackage{graphicx}	
\usepackage{amsmath}	
\usepackage{amssymb}	
\usepackage{multicol}        
\usepackage{bm}		
\usepackage{pdflscape}	

\DeclareRobustCommand{\VAN}[3]{#2}
\let\VANthebibliography\thebibliography
\def\thebibliography{\DeclareRobustCommand{\VAN}[3]{##3}\VANthebibliography}

\usepackage{graphicx}
\usepackage{float}

\usepackage{tabularx}
\usepackage{color}
\usepackage{url}

\usepackage{enumitem}
\usepackage{booktabs}
\usepackage{multirow}
\usepackage{subcaption}

\usepackage[T1]{fontenc}
\usepackage{ae,aecompl}

\usepackage{newtxtext,newtxmath}

\title[Kinematic satellite planes in TNG50]{A statistical look on kinematic planes of satellite galaxies I: frequency and properties  in TNG50 MW/M31-like galaxies}

\author[G\'amez-Mar\'in et al.]
{Mat\'ias G\'amez-Mar\'in,$^{1}$\thanks{E-mail: matias.gamez@estudiante.uam.es}
Rosa Dom\'inguez-Tenreiro,$^{1,2}$
Isabel Santos-Santos,$^{3}$
\newauthor
Susana E. Pedrosa$^{4}$
\\
$^{1}$Departamento de F\'isica Te\'orica, Universidad Aut\'onoma de Madrid, E-28049 Cantoblanco, Madrid, Spain \\
$^{2}$Centro de Investigaci\'on Avanzada en F\'isica Fundamental, Universidad Aut\'onoma de Madrid, E-28049 Cantoblanco, Madrid, Spain \\
$^{3}$Institute for Computational Cosmology, Department of Physics, Durham University, South Road, Durham, DH1 3LE, UK
\\
$^{4}$Instituto de Astronom\'ia y F\'isica del Espacio, CONICET-UBA, 1428, Buenos Aires, Argentina}

\begin{document}
\label{firstpage}
\pagerange{\pageref{firstpage}--\pageref{lastpage}}
\maketitle

\begin{abstract}

We use the TNG50 simulation to explore the possible existence of satellite galaxy sets, with fixed-in-time identities, forming kinematically-persistent planes (KPPs) along cosmic time around 190 MW/M31-like galaxies. 
This is the first study to assess their frequency within the $\Lambda$CDM framework. 
We identify KPPs around 46 of these host galaxies, with at least 25\% of their satellites in such configurations. 
Thereby, KPPs appear more frequent than previously reported, 
appearing in $\sim24\%$ of MW/M31-like systems, and in $\sim40\%$ of those populated with $N_{\rm sat}\geq9$.
We find a dependency of the former frequency on the minimum satellite stellar mass cut, suggesting that it would increase with higher mass resolution.
KPP satellite members form a distinct set compared to satellites outside KPPs, located at further distances from the center of their host and maintaining higher specific angular momentum since high redshift.
KPP satellites form thin and oblate planes in positional space during long periods of cosmic time.
We statistically confirm that KPPs form a kind of backbone of observationally-detected positional planes, and that, in velocity space, KPPs behave as kinematic morphological disks.
We show that KPP formation, defined as the time when satellite orbital poles align around a specific, fixed direction (occurring at Universe age $\sim4$ Gyr), predates the end of halo's fast-phase of mass assembly, indicating that halo processes do not drive this clustering. 
Finally, our results are broadly consistent with the MW's kinematic plane at $z=0$ concerning its morphological properties and degree of satellite orbital poles clustering, mitigating the tension between the existence of these structures and the $\Lambda$CDM paradigm.

\end{abstract}



\section{Introduction}
\label{ch4:sec:intro}
The accepted cosmological paradigm, known as the $\Lambda$CDM model, has been succesful at describing how the formation and evolution of cosmic structures across a wide range of scales, from large-scale cosmic structures up to clusters and galaxies. However, some challenges arise when considering smaller cosmological scales, particularly concerning  dwarf galaxies. 
One notable issue that $\Lambda$CDM encounters is the satellite plane problem. This problem emerged with the discovery that satellites around the MW exhibit a highly anisotropic distribution, forming a plane perpendicular to the galactic disks, a phenomenon first noted several decades ago \citep{Lynden76,Kunkel76}. The subsequent discovery of fainter satellites around the MW has only increased the fraction of satellites belonging to the plane. Additionally, a significant fraction of satellites forming this plane show kinematical coherence within the plane, see e.g. \citet{SantosSantos2020I}, hereafter Paper I,  \citet{Li21} and \citet{Taibi24}, suggesting that at least this fraction of satellites is rotating within the plane, contributing to its stability over time.

Similar phase-space correlations among satellites have also been observed in  M31 \citep{Koch06,McConnachie06,Metz07}, and other major galaxies outside the Local Group \citep[e.g.,][]{Chiboucas13,Tully15,Muller17,Muller18,Heesters21,MartinezDelgado21,Paudel21,Karachentsev24}.

These observations have sparked debates regarding the formation and stability of satellite positional planes
as well as their expectation within a cosmological context. Different studies have attempted to detect similar satellite  phase-space configurations and assess their frequency in $\Lambda$CDM simulations. While positional planes can be found, their presence was found to be  quite rare in the simulations in use and often transient, with the membership of at least a fraction of satellites being lost within short timescales \citep{Libeskind05,Libeskind09,Lovell11,Bahl14,Cautun15,Buck16,Ahmed17,SantosSantos2020I}. However, some positional planes exhibit kinematic coherence \citep[e.g.,][]{Shao19,Samuel2021,Santos-Santos_2023,Gamez-Marin2024}, although most of them are unstable and short-lived \citep{Bahl14,Gillet15,Buck16,Maji17b,ZhaoXinghai2023,Xu23}.

In line with these findings, \citet[][ hereafter Paper II]{SantosSantos2020II}, in their analysis of two hydrodynamical, zoom-in MW-mass systems, found positional planes and  observed important changes in their properties  over time. Although this suggests that transient satellites contribute to the apparent structure of these planes, it remains possible that positional planes may be endowed with a kind of persistent kinematic skeleton, responsible for their long-term stability \citep[see also][]{Gillet15}.
Indeed, at each simulation time step, they observed that a fraction of those satellites were in coherent co-orbitation\footnote{As in Paper I, II and III, in this Paper the term  co-orbitation will mean kinematic coherence, \textit{no matter the sense of rotation}, within an aperture $\alpha_{\rm co-orbit} = 36.^\circ87$, see \citet{Fritz18}.} within the planes. 

Several studies have focussed  directly on the study of kinematic-coherent planes, analyzing various properties to study their formation processes. \citet{Garaldi18} studied the clustering of satellites orbital poles 
(i.e., of the unitary vectors in the direction of satellite  angular momentum vectors,  $ \vec{J}^{\rm sat}_{\rm orb}/|\vec{J}^{\rm sat}_{\rm orb}|$) in 4 galactic systems.
Interestingly, they detected satellites with their orbital poles clustered along a specific direction in two of these galaxies. They concluded that, although their assembly histories do not determine the presence of planar structures, 
the assembly histories of the systems affected the formation timescales of the kinematic planes.
Other studies have aimed to replicate the specific kinematical coherence observed among the MW  'classical' satellites, where 8 out of 11 satellites are co-rotating. \citet{Shao19} find that a $30\%$  of all thin planes with 11 members in the EAGLE-100 volume exhibit such coherence, with orbital poles more clustered at lower redshifts. This suggests that kinematic structures like that of the MW may be dynamically young and have formed recently.
Along these lines, \citet{Sawala22} concluded, based on satellite orbit integrations, that the MW's classical satellite plane is a momentarily and short-lived structure \citep[see also][]{Maji17a,Lipnicky17}.
%

The role that kinematic support plays at ensuring the stability and long-term persistence of such satellite  planes, and  the age at which a system acquires its co-orbitation coherence, is still under debate.
\citet{Santos-Santos_2023}, hereafter Paper III \citep[see also][Paper IV]{Gamez-Marin2024}, addressed this issue through a systematical and detailed study of two zoom-in hydrodynamic simulations in a cosmological context, analyzing the same host-satellite systems as the ones analyzed in Paper II. They extended their analysis to the six-dimensional space, exploring the conservation of satellites' orbital angular momentum over long time intervals, and the clustering of their orbital poles along a direction of maximum co-orbitation, fixed across time. 
These authors identified the so-called Kinematically-Persistent Planes (KPPs) of satellites: 
sets of satellites whose identity is the same along evolution, and whose orbital poles are clustered around a fixed-in-time direction over long time intervals, thereby forming a kinematic structure of satellites that persists in time.
It is worth noting that  the positions of KPP satellite members vary with time, but the fitted plane to these positions (i.e., the so-called kinematic plane) is thin and oblate during long periods of time. These authors also found that  
the best fitting to positionally-identified planes (in Paper II) coincide with the detected kinematic planes (identified in Paper III, see fig.~7), except for a fraction of transient satellites that temporarily share the same spatial configuration but that are shortly replaced by other transient satellites. 

%
%
%
%
Despite the differences between the two zoom-in simulations used in these papers (in terms of the hydrodynamics implemented, subgrid physics, etc.), analyzing such low number of simulations is not statistically sufficient to draw robust conclusions about the KPP issue. 
In this sense, extending the work done in Papers III and IV to large-volume cosmological simulations is necessary.
These larger simulations provide with a significant range of density environments around host-satellite systems, allowing for the analysis of different scenarios in which KPPs are likely to form. Additionally, they enable obtaining a statistically significant fraction of KPPs, which is crucial for quantifying the frequency of KPPs around host-satellite systems.

This paper is the first in a series of two papers aimed at detecting early KPPs in the large-volume and high-resolution TNG50-1 cosmological simulation. Specifically, in this paper we will apply the methodology presented in Papers III and IV for detecting KPPs around MW-like disk galaxies, characterize their formation timescales, the orbital properties of satellites belonging to KPPs, and assess the possible effect of halo mass assembly history on KPP formation or destruction. The conclusions obtained in Paper IV regarding early mass flows around galaxy-to-be objects and their impact on KPP satellite dynamics will be tested in a forthcoming paper (G\'amez-Mar\'{i}n  et al., in prep.).

The paper is organized as follows: Sec.~\ref{ch4:sec:Simus} introduces the simulation analyzed and the host-satellite  systems, focusing on their selection criteria and their halo mass assembly histories. 
Sec.~\ref{ch4:sec:LookingKPPs} reports on the detection of KPPs in TNG50, whose co-orbitation quantification is analyzed in Sec.~\ref{ch4:sec:Quant_SatCoorb}. 
Sec.~\ref{ch4:sec:SatOrbits} presents an orbital analysis of the different satellite samples.
The properties of kinematically-persistent planes as positional planes are studied in Sec.~\ref{ch4:sec:KPP_PosSpace}, while in Sec.~\ref{ch4:sec:krot} we characterize satellites' motion within their corresponding KPP via the so-called `kinematic morphological' parameter \citep{Sales:2012}. An analysis of the distribution of the different timescales that arise from this study is presented in Sec.~\ref{ch4:sec:VariableRelationships}, along with possible correlations between them. The robustness of our results is tested in Sec.~\ref{ch4:sec:Discussion}, where we discuss whether satellites in KPPs are distinguishable groups of satellites among the total satellites population in terms of their masses, the possible bias introduced due to our satellites selection criteria, 
the resemblance of our KPPs to the MW's plane of satellites, and comparisons with previous works. Results are summarized in Sec.~\ref{ch4:sec:Summary}, along with the conclusions reached in this work. 
Finally, a glossary is provided to explain main terms and concepts used in this work.

\section{Simulation and host-satellite systems}
\label{ch4:sec:Simus}

\begin{figure*}
\centering
\includegraphics[width=0.47\linewidth]{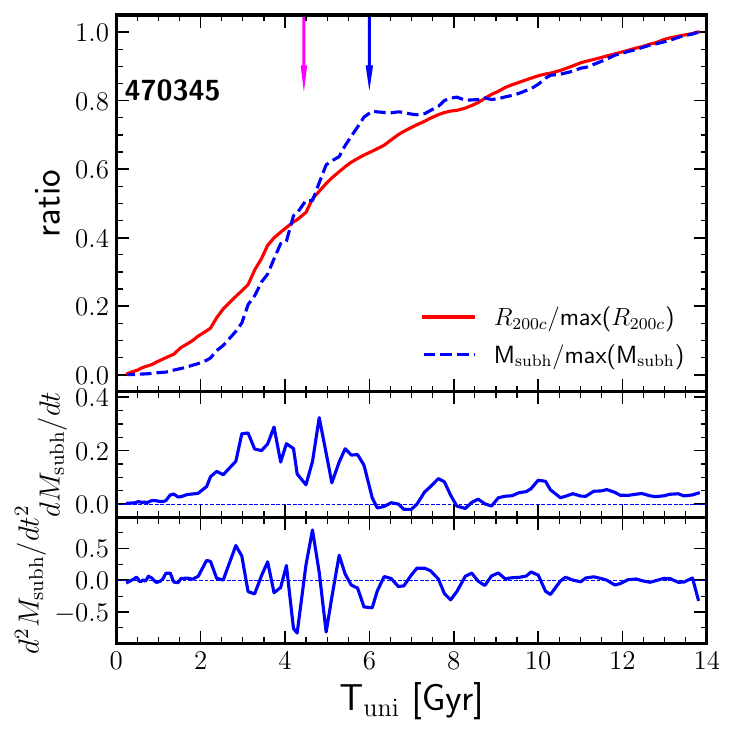}
\includegraphics[width=0.47\linewidth]{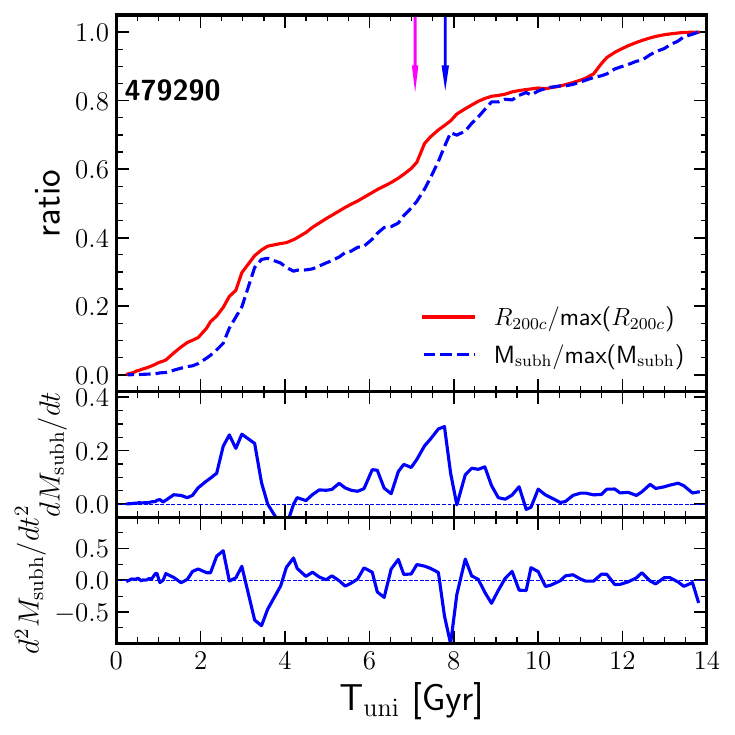}
\caption{Mass assembly histories and $R_{\rm 200c}$ evolutions for two HS systems, with ID\# given on the top left  corner of each plot (see Tab.~\ref{ch4:tab:DataTable1}).
Red lines correspond to the $R_{\rm 200c}(t)$ curve. Blue dashed lines stand for the mass of all bound particles inside $R_{\rm 200c}$, excluding subhalos other than the central one, $M_{\rm subh}(t)$.
Blue solid lines in the bottom panels depict the first and second order time derivatives, respectively,  of the $M_{\rm subh}(t)$ curve. Blue and magenta arrows mark the time $T_{\rm no-fast}$ and the time when the halo reaches half of its final $M_{\rm subh}$, $T_{\rm half,mass}$.
}
\label{ch4:fig:MassAssemHis}
\end{figure*}

\subsection{Illustris-TNG50}
\label{ch4:sec:TNG50}

In this study, we make use of one of IllustrisTNG simulations for galaxy formation \citep{Marinacci18,Naiman18,Nelson18,Pillepich18a,Springel18}. IllustrisTNG  assumes a cosmology consistent with the \citet{Planck16}: a flat $\Lambda$CDM cosmology with $H_{0}=67.74$ km$\cdot$s$^{-1}$Mpc$^{-1}$, $\Omega_{m}=0.3089$,  $\Omega_{b}=0.0486$, $\sigma_{8}=0.8159$, and $n_{s}=0.9667$.

We adopt the highest resolution flagship of IllustrisTNG, the TNG50-1 (hereafter TNG50) simulation \citep{Pillepich19,Nelson19b}. TNG50 is one of the few simulations able to resolve and follow the evolution of low-mass galaxies in a large cosmological context. This is a necessary condition to statistically study the origin and frequency of KPPs around central galaxies with a well-resolved and numerous satellite population. 
The TNG50 simulation evolves dark matter (DM), cosmic gas, stars, super massive black holes, and magnetic fields from $z=127$ to the present epoch, within a cosmological volume of $(\sim51.7$ cMpc)$^{3}$ box. The total run consists of 100 snapshots, typically spaced by 150 Myr in time. The particle mass resolution is $m_{\rm bar}=8.5\times10^{4}M_{\odot}$ for baryonic particles, and $m_{\rm DM}=4.6\times10^{5}M_{\odot}$ for dark matter.

TNG50 employs the moving mesh code AREPO \citep{Springel10AREPO} for simulating baryons, implementing prescriptions for star formation, stellar feedback, metal enrichment and black hole physics (seeding and feedback). For more detailed prescriptions of these implementations see \citet{Weinberger20} and references therein.
%
Furthermore, the TNG database provides tables of groups, subhalos, particles, subhalo merger trees, as well as other supplementary data catalogues. The groups are identified using the Friends-of-Friends algorithm on particles, while subhalos are structures within groups identified using the Subfind code \citep{SpringelWhite01}.

\subsection{Host Galaxies: the P24 sample}
\label{ch4:sec:Main}

We make use of the MW/M31-like galaxies presented in \citet{Pillepich24_MWM31} and study their satellite populations. We shall refer to these systems as host-satellite or HS systems. These host galaxies were selected after imposing several conditions that had to be simultaneously met, ensuring that their observational properties closely resemble those of the MW or M31:
\begin{enumerate}[label = {\roman*)}, wide, left=0pt, labelsep=0em,labelindent=0pt]
    \item Galaxies' stellar masses within 30 kpc from their center must be in the range $M_{\ast} = 10^{10.5}-10^{11.2}$ M$_{\odot}$.
    \item Galaxies must have a disk-like stellar morphology, constraining the values of the minor-to-major axis ratio of the stellar component distribution, and show observational evidence of spiral arms.
   \item An isolation criterion is required, such that galaxies must have no neighbouring galaxy with a stellar mass of $M_{\ast}\geq10^{10.5}$ M$_{\odot}$ within a 500 kpc distance at $z=0$. Additionally, they also impose an upper limit to the host halo mass, limiting it to values $M_{\rm 200c}<10^{13}$M$_{\odot}$. 
   This helps to exclude systems that have undergone extreme mergers at low redshift.
\end{enumerate}

These conditions return a total of 198 MW/M31-like galaxies, 190 of them being central galaxies of their FoF host halos, and 8 of them being FoF satellites of a more massive galaxy. We exclude this latter subgroup, keeping a sample of 190 MW/M31-like galaxies, hereafter the P24 sample. In a follow-up paper (G\'amez-Mar\'in et al. in prep.) we aim at characterizing the local Cosmic Web evolution surrounding these systems and relate it to the formation of KPPs. Such characterization would be strongly affected by the presence of a more massive companion.

Examples of the mass assembly history for sample halos are given in Fig.~\ref{ch4:fig:MassAssemHis}.
In these plots, red lines represent the time evolution of the radius of the sphere whose mean density is 200 times the critical density of the Universe, $R_{\rm 200c}$. Blue dashed lines correspond to the mass of all particles gravitationally bound to the host halo, $M_{\rm subh}(t)$. 
These data about the mass assembly history have been derived from the TNG50 subhalo catalogs and FoF merger trees.
The blue solid lines in the bottom panels depict the first and second-order time derivatives of the $M_{\rm subh}(t)$ curve, i.e., the rate of mass assembly and its variability, respectively. As such, the second derivative shows important fluctuations when fast rate changes occur. This is the case when two halos (or more than two, even a bound group of less massive halos) suffer from a fusion event (the final stage of merger events). Thus, fluctuations in the $d^2M_{\rm subh}(t)/ dt^2$ curve trace fusion events, as well as host halo dynamical activity, as we will see and quantify in Sec.~\ref{ch4:sec:Difs_KPP_nonKPP} and middle panels of Fig.~\ref{ch4:fig:Differences_HSsystems}.

The general trend illustrated in Fig.~\ref{ch4:fig:MassAssemHis}  reveals that the mass assembly rate (i.e., the slope of the $M_{\rm subh}(t)$ curve) is, in most cases, notably faster at earlier times compared to later epochs (see the lines on bottom panels). This is consistent with the two-phase mass assembly scenario for halos (and galaxies), characterized by an initial fast phase of high mass growth rates driven by merger activity within a local collapsing environment, followed by a subsequent slow phase marked by a decrease in 
halo assembly  activity. The two phases scenario was first established
through analytical models and N-body simulations \citep[see, for example,][]{Wechsler:2002,Zhao:2003,Brook:2005,SalvadorSole:2005,Griffen:2016}.
Later on, cosmological  hydrodynamical simulations \citep[e.g.,][]{DT:2006,Oser:2010,DT:2011} confirmed this scenario, as well as some of its implications on thick disk formation \citep{Brook:2004}, local elliptical galaxies properties \citep{Cook:2009}, and classical bulges formation \citep{Obreja:2013}, among other processes involved in  halo/galaxy mass assembly.

As explained in \citet{Padmanabhan93}, halo turn-around  leads to its decoupling from global expansion. Another important timescale refers to the rate of mass assembly stabilization (hereafter $T_{\rm no-fast}$), i.e., the end of the violent, fast phase of mass assembly, which can be either a distinct point in time or a time interval.
Indeed, in some cases, the slope of the $M_{\rm subh}(t)$ curve exhibits a distinct change, resulting in an isolated well-defined peak marked by a blue arrow in
 the $M_{\rm subh}(t)$ curve, as shown in the left panel of Fig.~\ref{ch4:fig:MassAssemHis}. 
This peak defines the Universe age  when the most  significant  decrease in the rate of mass assembly occurs
before becoming $\sim0$, when the first derivative (i.e. the rate of mass assembly) stabilizes.
This delineates a clear boundary between the fast and slow phases of mass assembly.
It is worth noting that, in some cases, this distinct peak  coincides with a major merger (fusion) event, as indicated by intense fluctuations in the $d^2M_{\rm subh}(t)/dt^2$ curve. 
Following this event, it is not uncommon that the $M_{\rm subh}(t)$  curves decreases towards increasing Universe age (T$_{\rm uni}$), taking some time for its first derivative to stabilize. 
In these cases, we will define $T_{\rm no-fast}$ as roughly the Universe age when this stabilization occurs.

In other cases, the change in the $M_{\rm subh}(t)$  slope is more gradual, with no well-defined, unique peak, as the right panel in Fig.~\ref{ch4:fig:MassAssemHis} illustrates. This situation is generally associated with  merging activity,  including the possible  accretion of  small halos over a broader time interval. This gradual change implies that these host halos are formed at later times compared to those that experience a well-defined fast phase period, i.e., they have an extended fast phase. 
See Sec.~\ref{ch4:sec:Difs_KPP_nonKPP} for the characterization of the dynamical activity through the standard deviation of the second derivative values.

\subsection{Satellite Samples}
\label{ch4:sec:Satellites}

\begin{figure}
\centering
\includegraphics[width=0.8
\linewidth]{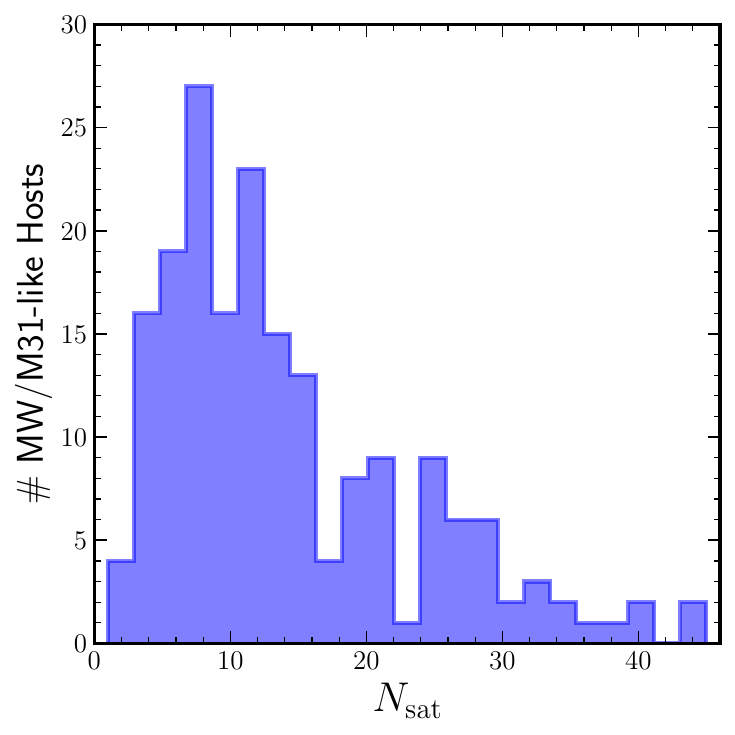}
\includegraphics[width=0.8\linewidth]{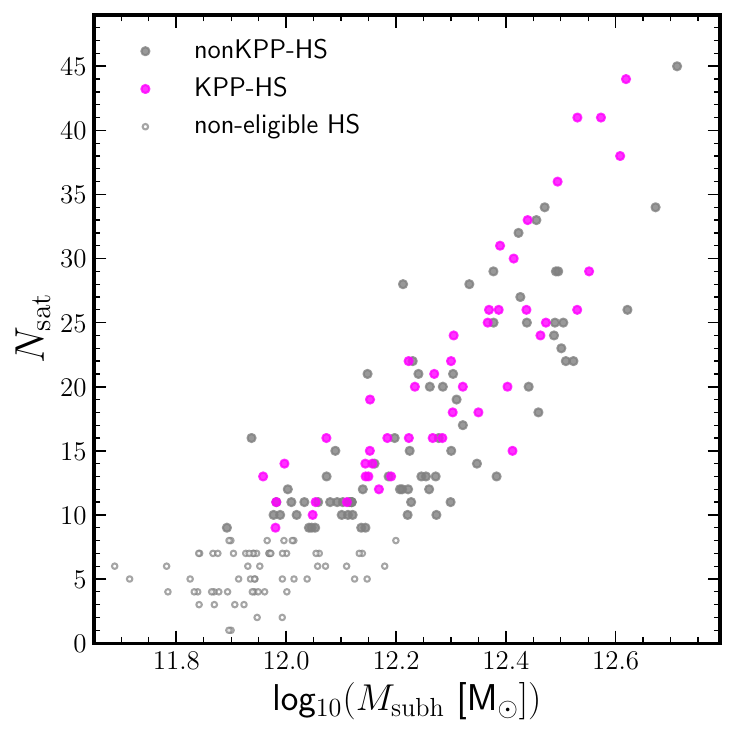}
\caption{Description of the P24 sample of MW/M31-like galaxy systems from \citet{Pillepich24_MWM31} and the final sample of host-satellite (HS) systems studied in this work. 
\textit{Upper panel}: distribution of the satellite number  per galaxy, $N_{\rm sat}$.
\textit{Lower panel}: the number of satellites per galaxy versus the corresponding host halo mass. Magenta (grey) filled points stand for those \textit{eligible} HS systems where KPPs have (have not) been identified, see Tab.~\ref{ch4:tab:DataTable1} and Sec.~\ref{ch4:sec:LookingKPPs}. Open grey circles represent those HS systems not suitable for our study as $N_{\rm sat}<9$ (see text for details).
}
\label{ch4:fig:SatNumber-Info}
\end{figure}

For each host galaxy, we consider as satellites candidates  those galaxies gravitationally-bound to the host halo at $z=0$, according to the Subfind halo finder.
To minimize numerical resolution effects, we apply a mass cut of a minimum of 10 gravitationally bound  stellar particles to each satellite, which translates in a satellite stellar mass lower-limit of $M_{\rm \ast,cut}=8.5 \times 10^5$ M$_{\odot}$.

%
In this way, the low-mass end of the collapsed objects mass function contributes a large number of low-mass objects compared to other large volume simulations.

To confirm that a given galaxy is a satellite, we follow back in time its orbit from $z=0$ to high redshift. We demand that satellites have infall into the host halo at least 500 Myr prior to T$_{\rm uni}(z=0)$. Backsplash galaxies are included in the sample if they remain bound at $z=0$ as well.

According to the selection criteria described above, the average number of satellites per host galaxy in the P24 sample  is $N_{\rm sat, aver}$ = 11$^{+9}_{-4}$, a rather high dispersion extending up to more than 40 satellites. This can be seen in the distribution of $N_{\rm sat}$ per system in the upper panel of Fig.~\ref{ch4:fig:SatNumber-Info}. 
%
%
In the lower panel of Fig.~\ref{ch4:fig:SatNumber-Info}, we analyze the possible  host halo mass effects on the number of satellites per host galaxy.
As expected, more massive host halos generally have more satellites. We will return to this plot in Sec.~\ref{ch4:sec:Freq_KPPHS}, where we study the frequency for HS systems of a given mass/number of satellites to host KPPs.

\section{KPPs in the P24 host-satellite systems}
\label{ch4:sec:LookingKPPs}

As explained in Sec.~\ref{ch4:sec:intro}, Kinematically-Persistent Planes (KPPs) of satellitesare fixed sets of satellites co-orbiting around their host galaxy, in such a way that their orbital poles are clustered along a specific direction and conserved across long cosmic time intervals. These structures are thought to be a kind of `backbone' of thin positional planes of satellites ensuring their long-term durability. Indeed, positional planes may persist because they consist of a skeleton of kinematically-coherent satellites plus interloper satellites. At given times, the latter may cross these planes, thus contributing to the population of the positional plane structure and enhancing its geometric properties. It is worth noting that some of these interlopers may show temporal kinematic coherence with the KPP satellite population as well, thus improving the plane's kinematic properties.

To identify a KPP in an HS system, we need to identify if a significant number of satellites are co-orbiting along a specific direction.
In order to determine the axis of maximum satellite co-orbitation ($\vec{J}_{\rm stack}$) in an HS system, we apply the same method as the one used in \citet{Santos-Santos_2023}, the so-called `Scanning of Stacked Orbital Poles Method', or $\vec{J}_{\rm stack}$ method. 
Essentially, $\vec{J}_{\rm stack}$ defines a direction, fixed in time, around which a maximum number of satellite orbital poles in the system cluster over time. Specifically, we use an angular aperture of $\alpha_{\rm co-orbit}=36.^{\circ}87$ to define co-orbitation of orbital poles \citep{Fritz18}\footnote{No distinction is made between those satellites rotating in the same and those rotating  in the opposite sense.}. KPP satellite members are then identified through an analysis of their pole conservation and clustering in the cosmic time interval from T$_{\rm uni}\simeq 6$ Gyr (a typical value for $T_{\rm no-fast}$, the halo mass stabilization timescale, see Tab.~\ref{ch4:tab:DataTable1}) to $z=0$.

\subsection{KPP selection criteria}
\label{ch4:sec:TNG50-KPPS}



We have applied the $\vec{J}_{\rm stack}$ method to the satellite samples of all the 190 HS systems in the P24 sample  (see Sec.~\ref{ch4:sec:Main}).
In most of the HS systems analyzed, we find satellites that conserve angular momentum during the slow phase of their host mass assembly period 
(see Sec.~\ref{ch4:sec:SatOrbPro} below; see also  fig.~6 in Paper III, and fig.~1 and sec.~3.2 in Paper IV for a detailed discussion of this behavior).
Moreover, in some HS systems, some satellites have their orbital poles clustered along a fixed common direction during long time intervals, defining a KPP.

In this work we shall focus on ``significant'' KPPs,  identified by satisfying specific selection criteria. Essentially, we keep those KPPs that have more than $N_{\rm KPP}^{\rm min}$ = 5 satellites members, and where the fraction of satellites in KPPs 
($f_{\rm KPP} = \frac{N_{\rm KPP}}{N_{\rm sat}}$) represents at least a $25\%$ of the total satellite population.
Note that this implies that HS systems with a total number of satellites $N_{\rm sat}\leq20$ have, at least, 25\% of their satellites within their KPP. 
No KPP structures have been identified in HS systems with $N_{\rm sat}<9$ (representing the 35 percentile of the distribution of $N_{\rm sat}$ around all hosts in the P24 sample, see top panel of Fig.~\ref{ch4:fig:SatNumber-Info}). We thus consider such low-$N_{\rm sat}$ systems (making a total of 67 HS systems in the P24 sample marked as open circles in Fig.~\ref{ch4:fig:SatNumber-Info}) as \textit{non-eligible} for our study on ``significant'' KPPs, and exclude them from further analysis.
The rest of HS systems (123) with $N_{\rm sat}\geq9$ will be referred to as the \textit{eligible} sample.

All HS systems hosting KPPs, hereafter KPP-HS systems, are listed in Tab.~\ref{ch4:tab:DataTable1}, where we give their halo identity numbers,  the total number of satellites identified,  $N_{\rm sat}$, and the fraction of them in KPPs, $f_{\rm KPP} = \frac{N_{\rm KPP}}{N_{\rm sat}}$, regardless of the sense of rotation. For the sake of completeness, we also give the satellite co-rotation fraction, $f_{\rm co-rot}$, as the larger fraction of satellites in a KPP rotating in the same sense with respect to $N_{\rm KPP}$.
Other parameters listed in this Table, characterizing different properties of KPP-HS systems,  will be defined along the next sections of this work. 

KPP-HS systems are ordered in Tab.~\ref{ch4:tab:DataTable1} roughly according to their satellite richness: the first block includes those KPP-HS systems with $f_{\rm KPP}>30\%$, the second block those with $25\% \leq f_{\rm KPP}\leq 30\%$, and the third block includes those systems with a low $N_{\rm sat}<20$ (note that these HS systems are biased against low $f_{\rm KPP}$ values).


\subsection{Frequency of KPP-HS systems in the P24 sample}
\label{ch4:sec:Freq_KPPHS}

After applying the previous conditions, we find a total of 46 KPP-HS systems.
These are marked as magenta points
in the bottom panel of Fig.~\ref{ch4:fig:SatNumber-Info}. On the other hand, we find 77 HS systems with $N_{\rm sat}\geq9$ in which no KPPs fulfilling the selection criteria have been detected, hereafter the nonKPP-HS sample. These are marked as grey filled points in the bottom panel of Fig.~\ref{ch4:fig:SatNumber-Info}.

The fraction of HS systems in the P24 sample where a KPP has been identified  relative to the 190 HS systems in P24 is $f_{\rm KPP-HS}^{\rm P24} = 46 / 190 \sim$ 24\%. 
It is worth noting that this fraction depends both on the KPP selection criteria, and on the satellite selection criteria used in the P24 sample (see Tab.~\ref{ch4:tab:FrequenciesTable}). 
In Sec.~\ref{ch4:sec:Disc_sel_criteria}, we vary each of these conditions individually and examine how the resulting frequencies change.

We now focus on the fraction of KPP-HS systems (magenta points in Fig.~\ref{ch4:fig:SatNumber-Info}) over the total number of \textit{eligible} HS systems (filled points in Fig.~\ref{ch4:fig:SatNumber-Info}), $f_{\rm KPP-HS}^{\rm P24-eli}$, which gives a value of $\sim37\%$. 
No clear dependence of this frequency with $M_{\rm subh}$ or the $N_{\rm sat}$ has been found in Fig.~\ref{ch4:fig:SatNumber-Info}, except maybe for the more  massive halo, more satellite-rich HS systems, where the statistics is poor and no robust conclusions can be reached.
Hence, we have found that a $\sim$ 40\% of HS systems in the P24 sample with a high enough $N_{\rm sat}$ that the search for KPPs can be undertaken,
do show a KPP orbiting around them, whatever their halo mass and the number of satellites.
It is worth mentioning that, contrary to the $f_{\rm KPP-HS}^{\rm P24}$ fraction, the value of $f_{\rm KPP-HS}^{\rm P24-eli}$ is not expected to show relevant changes with the satellite selection criteria (or, in a similar manner, with the mass resolution of the simulation), shall one keep fixed the KPP selection criteria, see Sec.~\ref{ch4:sec:Disc_sel_criteria}. 

The results presented here represent an important step forward in the statistical study of kinematic planes around MW/M31-like galaxies in $\Lambda$CDM simulations.  Importantly, our findings suggest KPPs might be more frequent than previously reported. 

\begin{figure*}
\centering
\includegraphics[height=6cm]{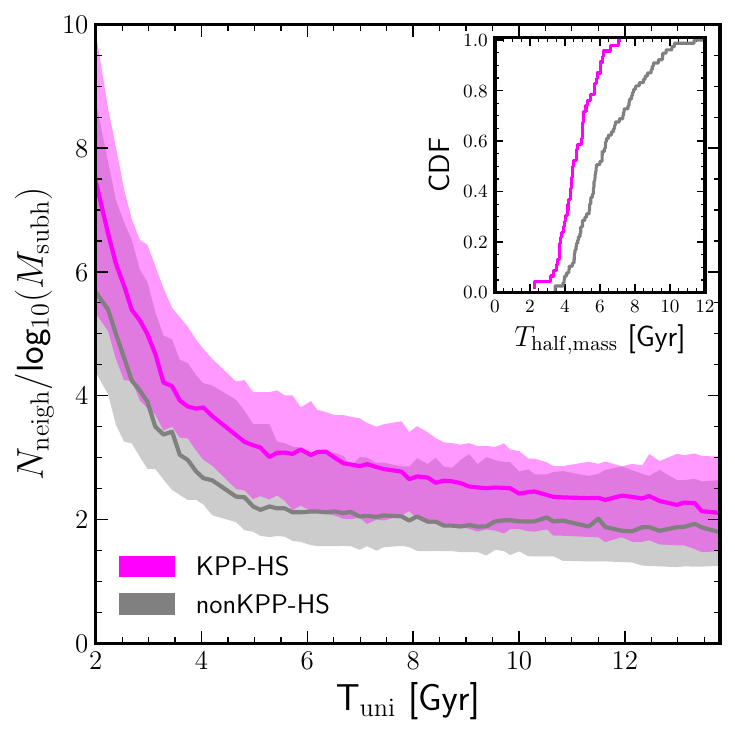}
\includegraphics[height=6cm]{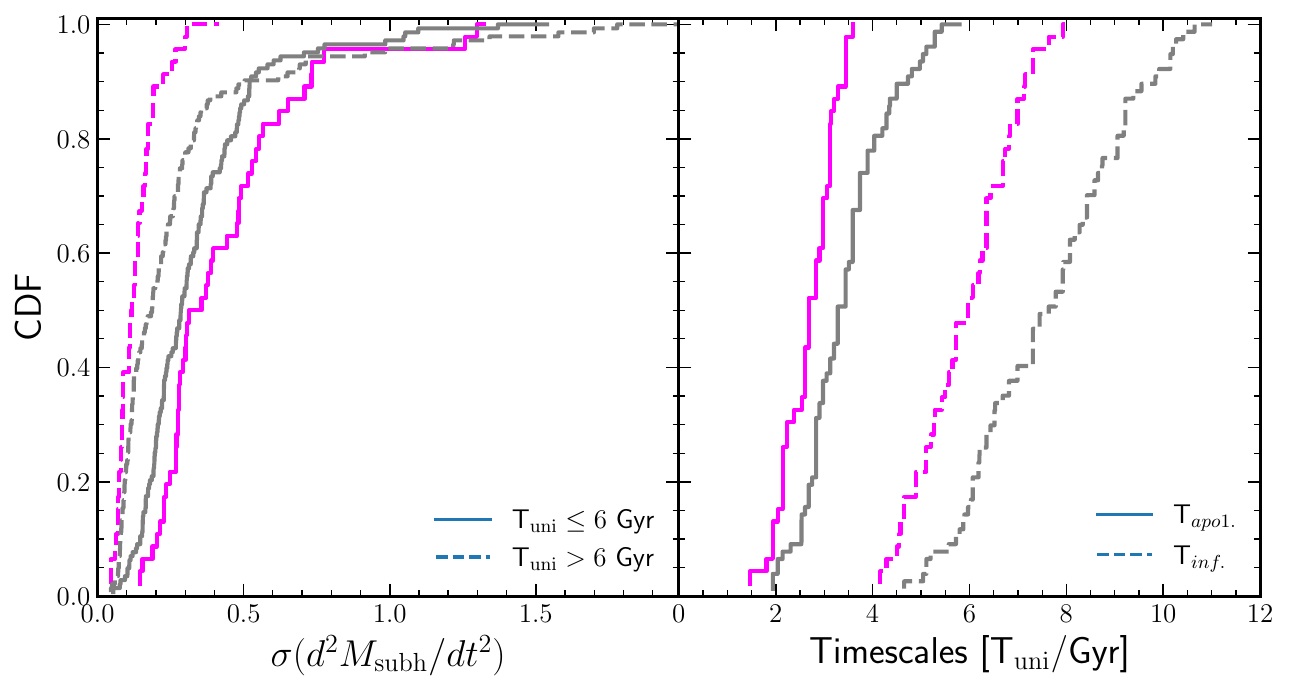}
\caption{\textit{Left panel}: Distribution of the number of neighbours, $N_{\rm neigh}$, within  $R_{\rm env}=500$ kpc from the halo center, for halos in the KPP-HS system sample (magenta lines), and halos in the nonKPP-HS system sample (grey lines) at different Universe ages T$_{\rm uni}$. 
The cumulative distribution of $T_{\rm half, mass}$ for the halos of all the HS systems studied in this work is depicted as an inset.
\textit{Middle panel}: Cumulative distribution functions (CDFs) of the dispersion of the $d^2M_{\rm subh}/dt^2$ values for KPP-HS (magenta lines) and nonKPP-HS (grey lines) systems at times prior (dashed lines) and after (solid lines) the typical mass assembly stabilization time of KPP-HS systems.
\textit{Right panel}: CDFs of the median first apocenter ($T_{\rm apo1}$) and infall ($T_{\rm inf}$) timescales of satellites in KPP-HS (magenta line) and nonKPP-HS (grey line) systems.}

\label{ch4:fig:Differences_HSsystems}
\end{figure*}

\subsection{KPP-HS vs nonKPP-HS systems}
\label{ch4:sec:Difs_KPP_nonKPP}


In the inset plot in the left panel of Fig.~\ref{ch4:fig:Differences_HSsystems} we show  the cumulative distribution function (CDF) of the half-mass timescales $T_{\rm half,mass}$ for the halo mass $M_{\rm subh}$ assembly.
We observe that hosts from the nonKPP-HS sample (grey CDF)  have a mass assembly remarkably retarded relative to that of halos in the KPP-HS sample (magenta CDF).
This result suggests different environment conditions in each case.



To catch possible environmental density  effects, we have calculated the number of neighbours, $N_{\rm neigh}$, in the vicinity of each halo at different epochs, making use of all galaxies at least as massive as our satellite population ($M_{\rm \ast}\geq8.5\times10^{5}$ M$_{\odot}$), and within a radial distance of 
$R_{\rm env}=2\times R_{200c}(z=0)$, as tracers.  
These results are depicted in the left panel of Fig.~\ref{ch4:fig:Differences_HSsystems}, where environments around KPP-HS halos (magenta line) and nonKPP-HS halos (grey line) are compared.
To extract the $N_{\rm sat}$ dependence on the halo mass (as shown by the bottom panel in Fig.~\ref{ch4:fig:SatNumber-Info}), the $N_{\rm neigh}$ has been normalized by the $\log(M_{\rm subh})$ at each redshift.
We find that  KPP-HS systems develop (i.e., are situated at earlier times) in environments with a higher number of neighbours, indicating a higher underlying mass density, and, therefore, a faster rate of structure formation. In contrast, nonKPP-HS systems are located in less dense regions during early times. 
It is worth noting that by  $z=0$,  the differences among the number of neighbours surrounding hosts in both samples are less important. 
%
These trends persist when changing the environment radius $R_{\rm env}$ 
and when limiting the definition of $N_{\rm neigh}$ to massive subhalos ($M_{\rm subh}\geq 10^{11.6}$M$_{\odot}$, the lower limit of our HS systems mass) around our hosts at different redshifts.
We recall that the P24 sample applies an isolation criteria of no massive companions within 500 kpc at $z=0$, but this does not preclude the possibility of these hosts suffering from 
possible  dynamical activity due to massive close companions at earlier times. 

This late dynamical activity that nonKPP-HS host halos suffer compared to KPP-HS systems is evidenced in their merger histories too. 
This is manifested in their $d^{2}M_{\rm subh}/dt^2$ curves, as nonKPP-HS show higher fluctuations in these curves compared to KPP-HS systems at late times (see bottom subpanels of Fig.~\ref{ch4:fig:MassAssemHis} for two examples; note that no example of nonKPP-HS systems is provided there). In order to account for this difference, in the middle panel of Fig.~\ref{ch4:fig:Differences_HSsystems} we plot the CDFs of the standard deviation of $d^{2}M_{\rm subh}/dt^2$  for each KPP-HS (magenta lines) and nonKPP-HS systems (grey lines). We consider two time intervals, i.e., before and after the median Universe age value of $T_{\rm no-fast}$ for KPP-HS systems (T$_{\rm uni}\simeq 6$ Gyr). While the dynamical activity of KPP-HS systems takes place at early times, nonKPP-HS systems show high dispersion values of $d^2M_{\rm subh}/dt^2$ at late Universe ages (T$_{\rm uni}>6$ Gyr) compared to KPP-HS systems, as shown in the middle  panel of Fig.~\ref{ch4:fig:Differences_HSsystems}. In other words, nonKPP-HS systems extend their violent mass assembly phase beyond T$_{\rm uni}\simeq6$ Gyr.

To gain more insight on the differences among KPP-HS and nonKPP-HS systems,
we study some satellite orbital parameters, finding that satellites belonging to nonKPP-HS systems come from further away in terms of their first apocentric distance. 
It has also been found they have later infall and first apocenter (turn-around) timescales, as depicted in the right panel of Fig.~\ref{ch4:fig:Differences_HSsystems} (see definitions in Sec.~\ref{ch4:sec:SatRadDist}). This suggests  a later stabilization of the system. 
%

The physical mechanisms responsible for KPP formation, or their absence, in HS systems will be studied in Gámez-Marín et al. in prep.. In the next sections we will focus on KPP-HS systems and the characteristics of their KPPs.


\section{Quantifying satellite co-orbitation in KPP-HS systems}
\label{ch4:sec:Quant_SatCoorb}

We quantify the quality of, and delve  deeper into, the co-orbitation behavior of satellites in the KPP-HS system  sample. 
To put this study into context, we establish the statistical sample we study.
Firstly, to enlarge the sample size, we use, for each of the 46 KPP-HS systems in Tab.~\ref{ch4:tab:DataTable1}, their  corresponding versions occurring in  each of the 49 simulation outputs used in this work from T$_{\rm uni}\simeq$ 6 Gyr onward.
Secondly, it turns out that the only reliable data on satellite galaxy kinematics are those relative to the MW at $z=0$.    We note that these data do not allow us to distinguish whether a given satellite orbital pole is  persistently or temporarily kinematically aligned with the axis of the kinematic MW plane.  The data only provide with information about  the value of the angle the pole and the axis currently form at z=0.   
Therefore, we need to determine the set of co-orbiting satellites (CS sets) in P24, at each of the 49 outputs we consider, and for each KPP-HS system. A CS set  at a given T$_{\rm uni}=t$, in a given KPP-HS system, consists of all the satellites members that at this particular T$_{\rm uni}$ are co-orbiting, irrespective that they are persistently or just temporarily co-orbiting.  In other words, to each KPP corresponds, at each output time T$_{\rm uni}=t$, a CS$(t)$ set of $N_{\rm CS}(t)$ satellites, formed by the  KPP member satellites  and the temporarily co-orbiting satellites.
To make  statistical analyses  we  use the information contained in 2254 CS planes.  This number comes from considering, for each of the 46 KPPs listed in Tab.~\ref{ch4:tab:DataTable1},  their  corresponding  CS$(t)$ versions occurring in  each of the simulation outputs from T$_{\rm uni}\simeq$ 6 Gyr onward. This sample will be known as the \textit{statistical sample of CS sets}.

\begin{figure*}
\centering
\subfloat{\includegraphics[width=0.4\linewidth]{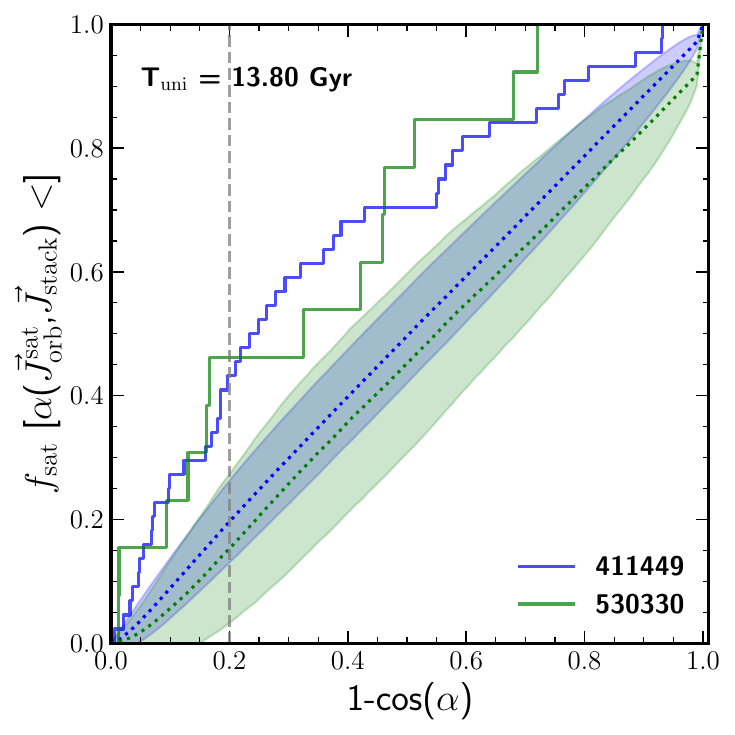}
{\hspace{0.1cm}(a)}}
\subfloat{\includegraphics[width=0.4\linewidth]{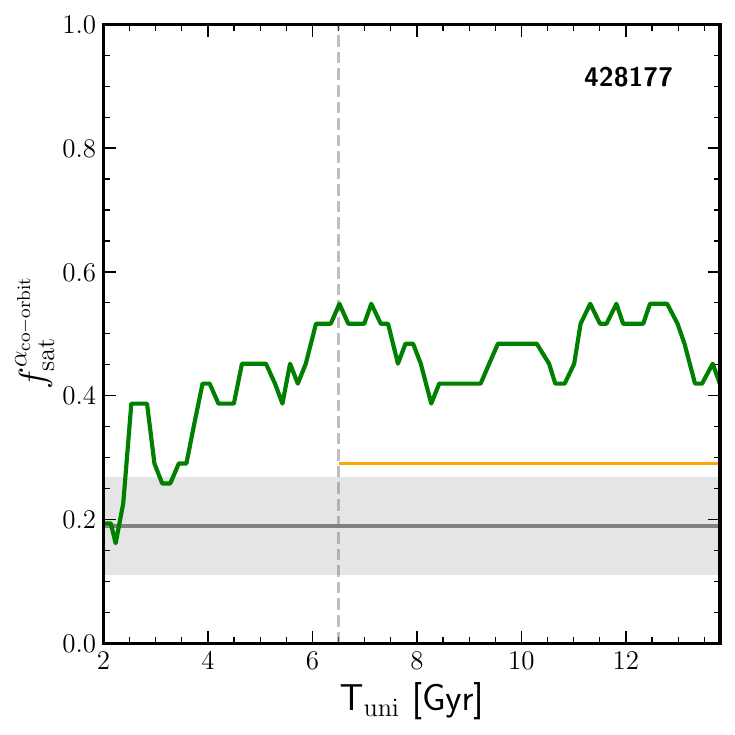}
{\hspace{0.1cm}(b)}}
\\
\subfloat{\includegraphics[width=0.4\linewidth]{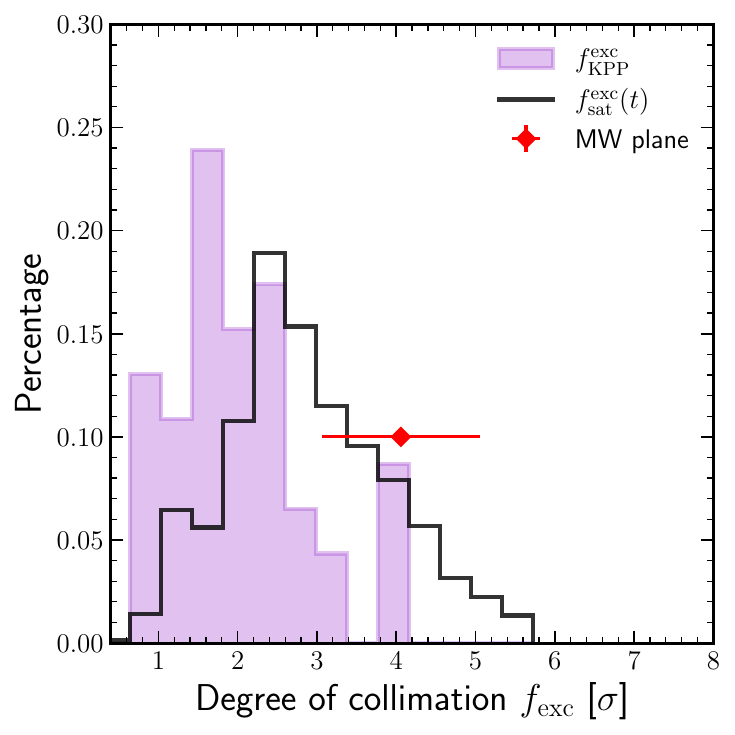}
{\hspace{0.1cm}(c)}}
\subfloat{\includegraphics[width=0.4\linewidth]{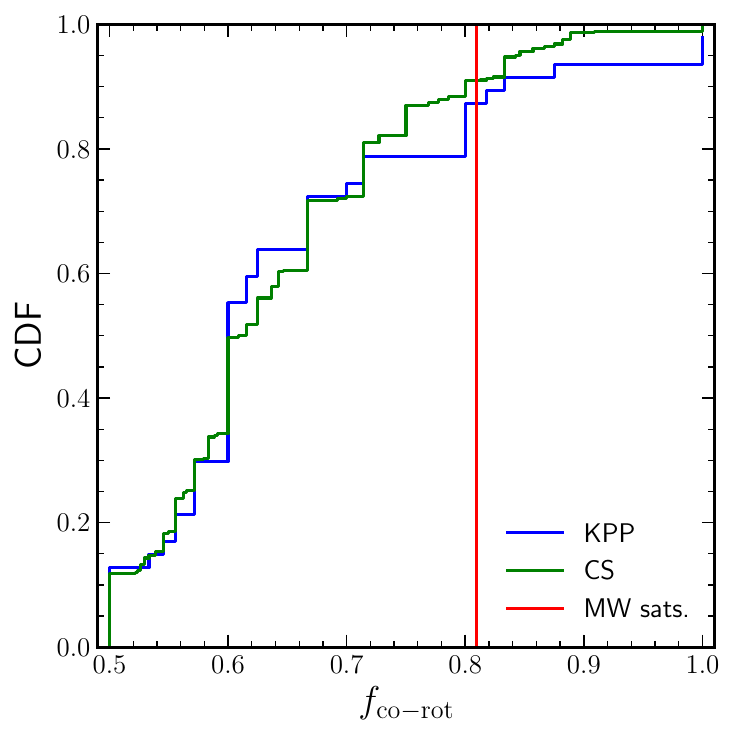}
{\hspace{0.1cm}(d)}}

\caption{
Analyses of the degree of satellite co-orbitation in KPP-HS systems. 
\textit{Panel (a)}: Co-orbitation of satellite orbital poles in two example KPP-HS systems 
(see legend and Tab.~\ref{ch4:tab:DataTable1}).
Solid lines show $f_{\rm sat}^{\alpha}(z=0)$, the cumulative fraction of satellites with orbital poles aligned within an angular distance $\alpha$ measured from the corresponding $\vec{J}_{\rm stack}$ axis, at $z=0$. 
Results for orbital pole randomized distributions for each of the analyzed KPP-HS systems are also shown as dotted lines (average values), with their corresponding  $\pm1\sigma$  highlighted by shaded areas. A vertical dashed line marks the angular distance $\alpha_{\rm co-orbit}=36.^{\circ}87$ that sets our criteria for co-orbitation.
\textit{Panel (b)}: $f_{\rm sat}^{\alpha}(t)$ curve at $\alpha_{\rm co-orbit}$ aperture (green line), $f_{\rm sat}^{\alpha_{\rm co-orbit}}(t)$.
Yellow line stands for the constant value $f_{\rm KPP}$ (see Tab.~\ref{ch4:tab:DataTable1}, third column). The black horizontal line with shades stands for the average ($f_{\rm rand}^{\alpha_{\rm co-orbit}}$) and dispersion ($\sigma$) for the corresponding  randomized distribution. 
\textit{Panel (c)}: Histograms of the $f_{\rm KPP}^{\rm exc}$  and $f_{\rm sat}^{\rm exc}(t)$ parameters for the KPP-HS systems sample,
see Eq.~\ref{ch4:eq:excess-def}.  
The latter histogram involves the 2254 planes of the statistical sample of CS sets.
A red diamond with error bars stands for the MW excess results (mean and dispersion), see Sec.~\ref{ch4:sec:Comparison_MW} for details.
\textit{Panel (d)}: CDF of the fraction of co-rotating satellites within KPPs ($f_{\rm co-rot}$) for KPP satellites (blue line) and the statistical sample of CS sets (green line) of all KPP-HS systems in Tab.~\ref{ch4:tab:DataTable1}.
A red vertical line marks the MW value as calculated from \citet{Taibi24} table 2 data.}

\label{ch4:fig:unocos}
\end{figure*}

Fig.~\ref{ch4:fig:unocos} (a) depicts, for two example KPP-HS systems, the $z=0$ CDF $f_{\rm sat}^{\alpha}(t)$ of satellites orbital pole angular distances relative to the axis of maximum co-orbitation,  $\vec{J}_{\rm stack}$. 
This $f_{\rm sat}^{\alpha}(t)$ is the fraction $N_{\rm sat}^{\alpha}(t)/N_{\rm sat}$, where $N_{\rm sat}^{\alpha}(t)$ is the number of satellites, at Universe age T$_{\rm uni}=t$, with orbital poles $\vec{J}_{\rm orb}$ such that the angle ang$(\vec{J}_{\rm stack},\vec{J}_{\rm orb})<\alpha$. 
This analysis is done for all satellites of a CS$(t)$, irrespective of whether satellites belong to KPPs or not, since, as mentioned, at a specific T$_{\rm uni} = t$ transient satellites can have their orbital pole aligned with $\vec{J}_{\rm stack}$, thus contributing to the value of $f_{\rm sat}^{\alpha}(t)$.
Dotted lines show the mean $f_{\rm rand}^{\alpha}$ obtained when the satellite orbital pole configuration between T$_{\rm uni}\simeq6$ Gyr and T$_{\rm uni}(z=0)$ is randomized. Shaded bands show the $\pm1\sigma$ dispersion. For a detailed description of the methodology for pole randomization, see Appendix C in Paper III.
%


Fig.~\ref{ch4:fig:unocos} (a) illustrates an important result:
a significant fraction of the total satellite sample exhibit co-orbitation (defined within an angular distance to $\vec{J}_{\rm stack}$ being less than or equal to $\alpha = \alpha_{\rm co-orbit}=36.^{\circ}87$, see vertical dashed line). A similar co-orbitation signal is found in all KPP-HS systems listed in Tab.~\ref{ch4:tab:DataTable1}.  As previously mentioned, computing the pole co-orbitation $f_{\rm sat}^{\alpha_{\rm co-orbit}}(t)$ (i.e. the value of $f_{\rm sat}^{\alpha}$ evaluated at $\alpha_{\rm co-orbit}$ for all the CS sets)
generally implies that, at certain time intervals, $f_{\rm sat}^{\alpha_{\rm co-orbit}}(t)$ is even higher than the fraction of KPP satellites in the system, $f_{\rm KPP}$. This is illustrated in Fig.~\ref{ch4:fig:unocos} (b), where we display the evolution of $f_{\rm sat}^{\alpha_{\rm co-orbit}}(t)$ (blue line), and the constant value of $f_{\rm KPP}$ (horizontal orange line). 
For a given HS, KPPs are but subsamples of their CS sets at different T$_{\rm uni}$.

In order to quantify the degree to which $f_{\rm sat}^{\alpha_{\rm co-orbit}}(t)$ and $f_{\rm KPP}$ deviate from a randomized distribution at $\alpha_{\rm co-orbit}$, we have computed the `excess' of these values relative to the mean value $f_{\rm rand}^{\alpha_{\rm co-orbit}}$ of the randomized satellite orbital pole distributions, normalized to the corresponding dispersion $\sigma$. 
For each KPP-HS system, they are defined as:
\begin{equation}
\begin{aligned}
&f_{\rm sat}^{\rm exc}(t) = \frac{(f_{\rm sat}^{\alpha_{\rm co-orbit}}(t) - f_{\rm rand}^{\alpha_{\rm co-orbit}})}{\sigma} \\
&f_{\rm KPP}^{\rm exc} = \frac{(f_{\rm KPP} - f_{\rm rand}^{\alpha_{\rm co-orbit}})}{\sigma}
\label{ch4:eq:excess-def}
\end{aligned}
\end{equation}

These quantities are listed in Tab.~\ref{ch4:tab:DataTable1} for each KPP-HS system, and their distributions are drawn in {Fig.~\ref{ch4:fig:unocos} (c)}. Note that the $f_{\rm sat}^{\rm exc}(t)$ refers to the whole statistical sample of CS sets.
As expected, the $f_{\rm KPP}^{\rm exc}$ distribution exhibits slightly lower values than $f_{\rm sat}^{\rm exc}(t)$. 
In fact, CS sets include $19^{+16.2}_{-13.2}$\% (median and 25-75th percentiles) more satellites than KPP satellite groups.
Lastly, the distribution of values of $f_{\rm sat}^{\rm exc}(t)$ of co-orbiting satellites reach up to $\sim5.5\sigma$. 
This is expected as some KPP-HS systems host more co-orbiting satellites than KPP satellites at given Universe ages, increasing the orbital pole anisotropical distribution at specific times.

Finally, another important parameter is $f_{\rm co-rot}$, the fraction of KPP satellites (and of CS sets) orbiting in a given sense over the total number of satellites in a KPP (in a CS set). Values for KPPs are listed in Tab.~\ref{ch4:tab:DataTable1} and their distributions (CDFs) are depicted in Fig.~\ref{ch4:fig:unocos} (d), blue line. We observe that more than 50\% of KPP-HS systems show $f_{\rm co-rot}$ $\leq$ 0.6, indicating that, for half the KPP-HS systems in P24, KPP satellites have no preferential sense of co-rotation. On the other hand, around 20\% (30\%) of KPP-HS systems show high co-rotation values above 0.8 (0.7), indicating that their KPP satellites are preferentially co-rotating in one specific sense.
Concerning CS sets (green line), we observe that the distribution of values is similar to that of KPP satellites, with 10\% (30\%) of them showing co-rotation values above 0.8 (0.7).

\vspace{-0.5cm}




\section{Satellite orbital analysis}
\label{ch4:sec:SatOrbits}

\subsection{Radial distances of satellites to the host center}
\label{ch4:sec:SatRadDist}

Studying radial distances evolution is interesting in that it allows us to, e.g., analyze satellites turn-around and infall into-the-halo timescales.
An illustration of satellite distance $r(t)$ curves in one example KPP-HS system can be found in Fig.~\ref{ch4:fig:RadDistWithSubSat-Evol}. Specifically, we define the epoch at which the first apocenter event occurs for the $i$-th satellite, denoted as $T^{i}_{\rm apo1}$, representing its turn-around timescale. Another important timescale is the Universe age when the $i$-th satellite first crosses into the virial radius of its host halo, $T^{i}_{\rm inf}$. We define the $T_{\rm apo1}$ and $T_{\rm inf}$ timescales as the median values of $T_{\rm apo1}^i$ and $T_{\rm inf}^i$ for a given set of satellites.

\begin{figure}
\centering
\includegraphics[width=0.9\linewidth]{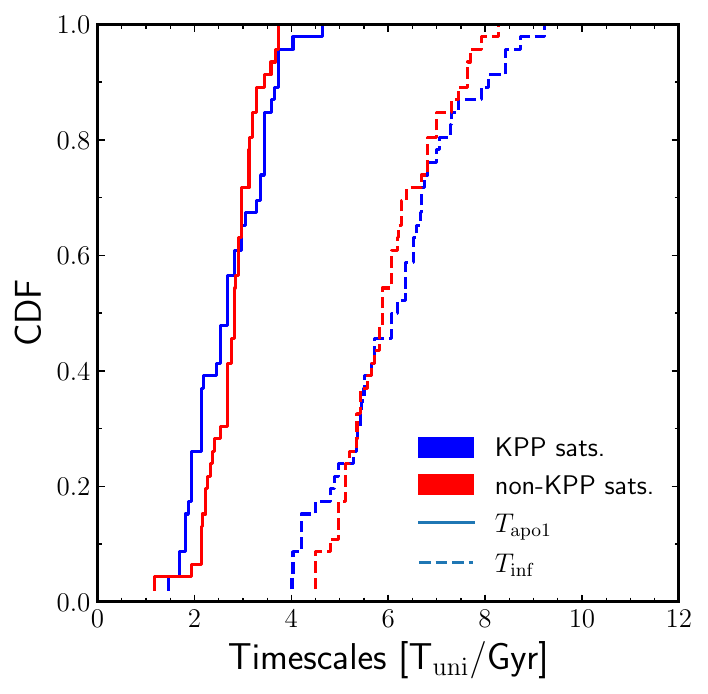}
\includegraphics[width=0.9\linewidth]{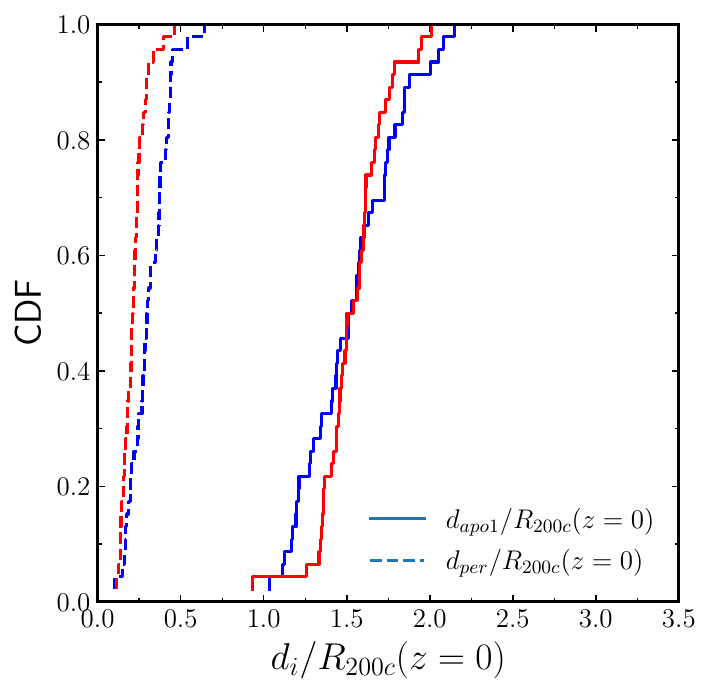}
\caption{Distribution of parameters concerning satellite orbits, for KPP (blue) and nonKPP member (red) satellites in KPP-HS systems.
\textit{Top panel}: CDF of the median values of satellite infall timescale, $T_{\rm inf}$, and the Universe age when each satellite reaches its first apocenter, $T_{\rm apo1}$, for each KPP-HS system.
\textit{Bottom panel}: CDFs for the normalized median pericentric ($d_{\rm per}(t)$/$R_{\rm 200c}(z=0)$) and first apocentric ($d_{\rm apo1}(t)$/$R_{\rm 200c}(z=0)$) distances.
}
\label{ch4:fig:DpNorm-Dist}
\end{figure}

The top panel in Fig.~\ref{ch4:fig:DpNorm-Dist} shows the CDF of the median infall and first apocenter timescales, $T_{\rm inf}$ and $T_{\rm apo1}$, for satellites belonging to each KPP-HS system, divided in KPP and non-KPP satellite populations. We see that the segregation between the distribution of different satellite populations in KPP-HS systems is not relevant. This suggests that these specific physical properties of satellites are not significantly influential at determining whether a given satellite belongs to a KPP or not.

Some other differences are apparent between the orbits corresponding to KPP and non-KPP satellites in KPP-HS systems, already shown for a given KPP-HS system in Fig.~\ref{ch4:fig:RadDistWithSubSat-Evol}. 
The bottom panel in Fig.~\ref{ch4:fig:DpNorm-Dist} shows the distribution of the median distances for the satellite sets belonging to the KPP-HS systems listed in Tab.~\ref{ch4:tab:DataTable1}. Specifically, we show the median pericentric distance normalized to the corresponding  $R_{\rm 200c}$ at $z=0$, $d_{\rm per}/R_{\rm 200c}(z=0)$.
A clear segregation stands out, with more than $90\%$ of non-KPP satellites (red lines) having their median pericenters between 0.1 and 0.3 of their $R_{\rm 200c}(z=0)$. In contrast, only 50\% of KPP satellites (blue lines) are within this interval, with most of them having larger pericenters within the $(0.15 - 0.45)\times R_{\rm 200c}(z=0)$ interval.
Thus, non-KPP satellites spend a fraction of their time in regions closer to the host center,
implying that they are likely to have experienced more frequent host-satellite and satellite-satellite interactions due to the higher number density of perturbers in those regions.
%
%
%

Regarding first apocentric distances, 
KPP-HS systems satellite distributions also reveal a difference between populations. 
%

\subsection{Satellite orbital angular momentum}
\label{ch4:sec:SatOrbPro}

The presence of KPPs in a given HS system is physically based upon orbital pole conservation of its satellite members (see Sec.~\ref{ch4:sec:TNG50-KPPS}), and their clustering along long periods of cosmic time. 
To study this, the angular momenta of each satellite identified in the 190 HS systems analyzed in this work has been calculated relative to the host galaxy center of mass (c.o.m.).
An example of these analyses can be found in figures 1 of Papers III and IV. We see in those figures that, while KPP satellites conserve their poles since early times (prior to $T_{\rm no-fast}$), this is not always the case for satellites outside KPPs.
%

\begin{figure}
\centering
\includegraphics[width=0.9\linewidth]{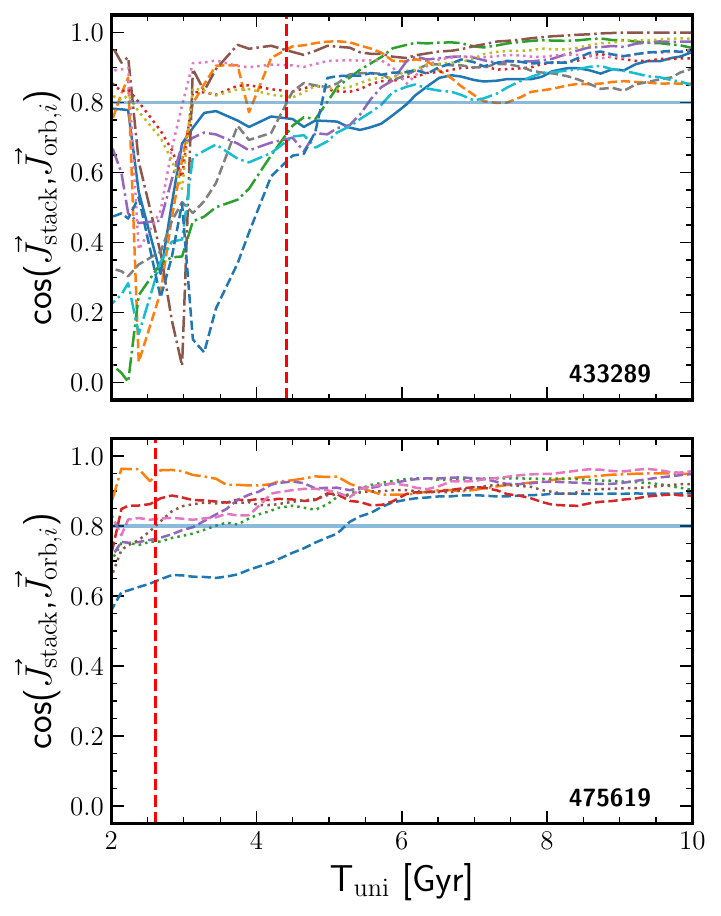}\\
\includegraphics[width=0.9\linewidth]{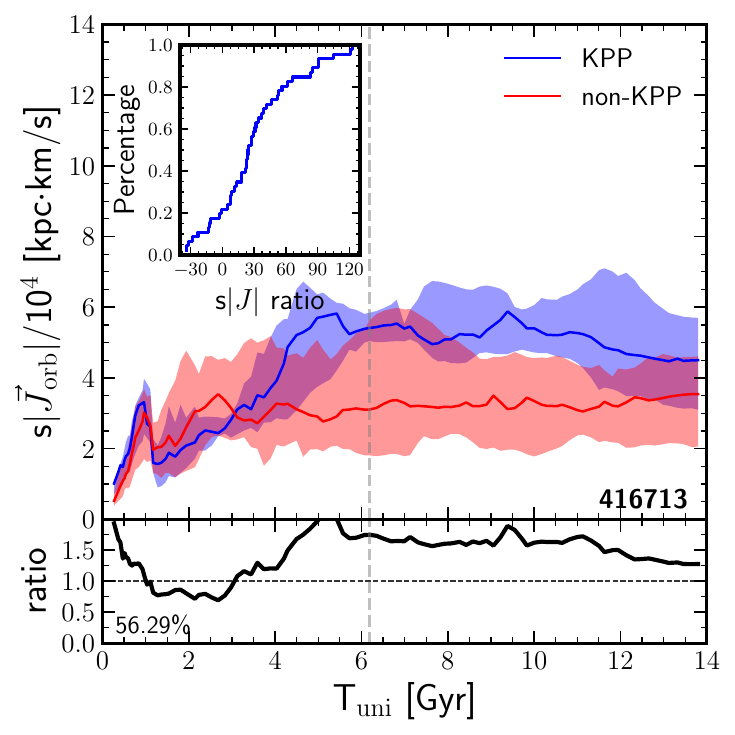}
\caption{\textit{Upper and middle panels}: Time evolution of the angles formed by the fixed $\vec{J}_{\rm stack}$ axial vectors  with the orbital poles of KPP satellites for two example KPP-HS systems (see Tab.~\ref{ch4:tab:DataTable1}). A horizontal line marks  the cosine of $\alpha_{\rm co-orbit}=36.^{\circ}87$. A red, vertical line marks the clustering timescale $T_{\rm cluster}^{\rm Jstack}$ for each of the two  KPP-HS systems.
\textit{Bottom panel}: 
Blue and red curves give the median values, at each time step, for the specific angular momentum of satellites that are KPP members and non-KPP members, respectively, for the \#433289 HS system. Shaded areas are the corresponding $25-75$th percentiles. The continuous black line shows the evolution of the ratio between the median values of s$|\vec{J}_{\rm orb}|$ for KPP satellites over that of non-KPP satellites at each time step. The dotted line (equality ratio = 1) serves as a visual guide. A vertical grey dashed line indicates $T_{\rm no-fast}$.
The CDF for the mean s$|J|$ratio (see Eq.~\ref{ch4:eq:sJratio}) corresponding to all the KPP-HS systems  is depicted as an  inset. 
}
\label{ch4:fig:Clus-SJ}
\end{figure}

Another way of analyzing orbital pole conservation of KPP satellites is illustrated in the upper and middle panels of Fig.~\ref{ch4:fig:Clus-SJ}. 
These panels display the time evolution of the angles formed by the respective fixed $\vec{J}_{\rm stack}$ axial vector  with the orbital poles of KPP satellites. It is evident that the orbital poles of individual satellites remain at a roughly constant angle from the $\vec{J}_{\rm stack}$  vector from a given Universe age onwards. 

Regarding the acquisition of orbital angular momentum by satellites, we have computed the specific angular momentum s$|\vec{J}_{\rm orb}^i|$ for individual satellites belonging to the KPP and non-KPP populations in KPP-HS systems, respectively.
The median s$|\vec{J}_{\rm orb}|$  and $25-75$th values at each timestep are represented in the bottom panel of  Fig.~\ref{ch4:fig:Clus-SJ} for an example KPP-HS system. 
%
%
%
%
Some  differences between KPP and non-KPP satellites stand out for most KPP-HS systems and are illustrated here.
Initially, KPP satellites show a rapid increase in their s$|\vec{J}_{\rm orb}|$, reaching relatively high median values at high redshift, in consistency with the angular momentum acquisition predicted by the so-called Tidal Torque Theory 
\citep[see, e.g.,][for a review]{Peebles:1969,Doroshkevich:1970,White:1984, Schafer:2009}.
Subsequently, this angular momentum acquisition is then progressively halted and, in some cases, it even decreases. Finally, at lower redshifts, median values of s$|\vec{J}_{\rm orb}|_{\rm KPP}$  are generally well conserved, except in instances of strong dynamical perturbations.

In contrast, the median s$|\vec{J}_{\rm orb}|_{\rm non-KPP}$  values are modest in many KPP-HS systems. 
%
To quantify the extent to which KPP satellites have higher median values of s$|\vec{J}_{\rm orb}|$ compared to those for non-KPP satellites, we compute the time-averaged deviation from equality of the ratio between the s$|\vec{J}_{\rm orb}|$ values of KPP satellites with respect to the non-KPP ones, from $T_{\rm no-fast}$ onwards (see vertical dashed line), which we define as the s$|J|$ ratio (\%):
\begin{equation}
    \mathrm{s}|J| \mathrm{ratio} = \frac{1}{\mathrm{T_{uni}}(z=0) - T_{\rm no-fast}} \int^{\mathrm{T_{uni}}(z=0)}_{T_{\rm no-fast}}{\left(\frac{\mathrm{s}|\vec{J}_{\mathrm{orb}}|_{\mathrm{KPP}}}{\mathrm{s}|\vec{J}_{\mathrm{orb}}|_{\mathrm{non-KPP}}}-1\right)dt}
\label{ch4:eq:sJratio}
\end{equation}

The CDF of these values for all KPP-HS systems is provided as an inset in Fig.~\ref{ch4:fig:Clus-SJ}. For more than an 80\% of the KPP-HS systems the average s$|J|$ ratio values are positive, corroborating that, generally, KPP satellites have higher specific angular momenta than non-KPP ones. 





\subsection{The clustering timescale}
\label{ch4:sec:Tcluster}




As illustrated in the top and middle panels of Fig.~\ref{ch4:fig:Clus-SJ}, the time evolution of the angles ang($\vec{J}_{\rm stack},\vec{J}_{\rm orb}$) for the two KPP-HS systems shown, indicates that KPP satellites have their orbital poles highly aligned from early Universe ages, maintaining this alignment from a given moment onwards.
%
%

A timescale for the establishment of orbital pole clustering (and hence the formation time of the KPP), $T_{\rm cluster}^{\rm Jstack}$,  can be defined for each KPP as the median value of the Universe age when the different KPP satellite members align with $\vec{J}_{\rm stack}$ (cos$(\vec{J}_{\rm stack},\vec{J}_{\mathrm{orb},i})\geq0.8$, corresponding to $\alpha\leq\alpha_{\rm co-orbit}$).
Top and middle  panels in Fig.~\ref{ch4:fig:Clus-SJ} show the respective $T_{\rm cluster}^{\rm Jstack}$ timescales as vertical red lines. The corresponding $T_{\rm cluster}^{\rm Jstack}$ values for all KPP-HS systems are given in Tab.~\ref{ch4:tab:DataTable1}, and their distribution is shown in Figs.~\ref{ch4:fig:CDF-timescales} and \ref{ch4:fig:Timescales_visual_distrib_ch4}.
%
This distribution peaks at around T$_{\rm uni}=4.0$ Gyr, proving that satellite orbital pole clustering occurs at earlier times than their respective $T_{\rm no-fast}$ timescale.
This result, initially found in the two zoom-in HS systems hosting KPPs studied in Paper IV, is now reinforced with a statistical sample of HS systems.
We will revisit this point in Sec.~\ref{ch4:sec:VariableRelationships}, where we will compare the different timescales of various events related to the properties of KPPs.



\begin{table*}
\caption{Table containing an overview of the properties presented in this paper regarding HS systems  where KPPs have been detected (KPP-HS systems). The information is categorized in terms of KPP richness, orbital pole clustering timescales, degree of clustering of orbital poles, satellite infall and first-apocenter timescales, specific orbital angular momentum ratio between KPP and non-KPP satellite populations, morphological $\kappa_{\rm rot}$ parameter information, and halo properties. Each category provides insights into the dynamical behavior, spatial distribution, and evolutionary processes of the satellite populations within KPPs. Timescales are given in Gyr and $M_{\rm subh}$ in units of M$_{\odot}$ (logarithmic scale). Numbers within brackets below the column headers indicate the subsection or equation where the corresponding header definition can be found.
}
\scriptsize
%
\vspace{0.4cm}
\hspace*{-0.4cm}
\begin{tabular}{|l | c  | c | c | c | c | c | c | c | c | c | c | c | c | c | c | c |}
\hline
\multicolumn{1}{|c|}{} & \multicolumn{3}{ c|}{KPP information} & \multicolumn{1}{ c|}{Clustering} & \multicolumn{3}{ c|}{Degree of pole clustering}   &  \multicolumn{3}{c|}{Trajectory properties} & \multicolumn{3}{c|}{$\kappa_{\rm rot}$  information} & \multicolumn{2}{c|}{Halo information} \\
\hline
\multicolumn{1}{|c|}{ Host ID \#} &   \multicolumn{1}{c|}{$N_{\rm sat}$ } & \multicolumn{1}{c|}{$f_{\rm KPP}$} & \multicolumn{1}{c|}{$f_{\rm co-rot}$} & \multicolumn{1}{c|}{$T_{\rm cluster}^{\rm Jstack}$} & \multicolumn{1}{c|}{$f_{\rm sat}^{\rm exc}(z=0)$} & \multicolumn{1}{c|}{$f_{\rm KPP}^{\rm exc}$} & \multicolumn{1}{c|}{$\sigma$} & $T_{\rm inf}$ & $T_{\rm apo1}$ &\multicolumn{1}{c|}{s$|J|$ rat. (\%)} &  \multicolumn{1}{c|}{$\kappa_{\rm rot}^{\rm eq}$} & \multicolumn{1}{c|}{$T_{\rm Krot}^{\rm eq}$} &   \multicolumn{1}{c|}{$T_{\rm Krot}^{\rm  0.5}$} & \multicolumn{1}{c|}{$T_{\rm no-fast}$} & \multicolumn{1}{c|}{$M_{\rm subh}$} \\

       & [\ref{ch4:sec:Freq_KPPHS}]& [\ref{ch4:sec:Freq_KPPHS}]& [\ref{ch4:sec:Freq_KPPHS}]& [\ref{ch4:sec:Tcluster}]& [Eq.~\ref{ch4:eq:excess-def}]& [Eq.~\ref{ch4:eq:excess-def}]& [\ref{ch4:sec:Quant_SatCoorb}]& [\ref{ch4:sec:SatRadDist}] & [\ref{ch4:sec:SatRadDist}] &[Eq.~\ref{ch4:eq:sJratio}]& [\ref{ch4:sec:krot}]& [\ref{ch4:sec:krot}]& [\ref{ch4:sec:krot}]& [\ref{ch4:sec:Main}]& [\ref{ch4:sec:TNG50}]\\ 
\hline
402555 & 25 & 0.32 & 5/8 & 3.27 & 3.22 & 1.47 & 0.091 & 4.65 & 2.14 & -11.12 & 0.65 & 4.94 & 4.78 & 4.8 & 12.47 \\
411449    & 44& 0.318 & 8/14 & 4.53& 3.52 & 1.81 & 0.066 & 6.58 & 2.68 &  27.2   & 0.77 & 7.26 & 4.91  & 6.8& 12.62\\
430864   & 26& 0.346 & 4/9 & 5.47& 3.29 & 1.94 & 0.085 & 6.69 & 3.44 & 46.8   & 0.75 & 7.76 & 5.57 & 7.3& 12.37\\
432106    & 33 & 0.303 & 8/10 & 3.93& 1.84& 1.44 & 0.077 & 7.29 & 3.37 & 21.8   & 0.65 & 7.52 & 5.89  & 5.5& 12.44\\
433289    & 36 & 0.305 & 6/11 & 4.42& 2.23 & 1.48 & 0.074 & 6.07 & 1.94 & 51.9   & 0.73 & 6.37 & 5.53  & 5.4& 12.49\\
436932 & 26 & 0.307 & 7/8 &  3.41& 2.39 &1.48 & 0.085 &5.35 & 1.94 & 37.2   & 0.82 & 4.96 & 3.09 & 5.5& 12.44\\
447914    & 29 & 0.448 & 8/13 & 3.49& 1.84 &3.08 & 0.083 & 6.35 & 3.44 & -2.8   & 0.56 & 8.58& 8.35 & 6& 12.55\\
468590    & 30 & 0.5 & 8/15 & 5.70& 4.31 &3.90 & 0.080 & 6.52 & 3.28 & 30.8   & 0.58 & 8.28 & 7.99 & 7& 12.41\\
470345    & 26 & 0.385 & 7/10 & 4.20& 2.39 &2.39 & 0.085 & 5.50 & 2.83 & 24.7   & 0.57 & 5.62 & 5.38 & 6& 12.39\\
471248  & 20 & 0.400 & 8/8 & 3.49& 2.71 &2.21& 0.101 & 5.44 & 1.87 & 24.2   & 0.72 & 6.05 & 4.28  & 6& 12.40\\
483594& 25 & 0.32 & 4/8 & 4.10& 2.34 & 1.47 & 0.091 & 5.57 & 2.34 & -33.5  & 0.61 & 6.08 & 4.28& 6.5& 12.37\\
488530& 22 & 0.318 & 5/7 & 2.36& 1.90 &1.43 &  0.096 & 6.19 & 1.69 & 17.9   & 0.84 & 5.10 & 3.38& 5.3& 12.22\\
489206  & 24 & 0.542 & 8/13 & 3.81& 4.30 &3.85& 0.092 & 6.07 & 2.14 & 7.5   & 0.67 & 6.09 & 4.33 & 5& 12.31\\
491426 & 23   &0.348  & 4/8 & 4.62& 3.04 & 2.07& 0.098 & 7.30 & 2.76 & 34.4   & 0.71 & 8.54 & 5.24& 4.5& 12.27\\
496186 & 22 & 0.455  & 6/10 & 3.95& 3.80 &2.85& 0.096 & 5.50 & 1.81 & 7.6   & 0.68 & 4.07 & 3.80 & 6& 12.30\\
501208& 20& 0.4    & 5/8 & 3.19& 2.71 & 2.21 & 0.101 & 4.89 & 2.14 & 23.0   & 0.75 & 4.46 & 3.33 & 6& 12.23\\

\hline
\hline
416713 & 38& 0.263& 5/10 & 3.56& 4.24 & 0.94 & 0.072 & 7.31 & 3.66 & 56.3  & 0.59 & 9.18 & 6.69& 7.3& 12.61\\
419618 & 41& 0.268& 9/11 & 4.51& 3.10 & 1.01 & 0.070 & 8.43 & 3.74 & 85.5  &0.78 & 8.78 & 6.0& 8.6& 12.53\\
421555 & 41& 0.268& 11/11 & 4.42& 3.45 & 1.01 & 0.070 & 7.45 & 3.44 & 66.5  &0.74 & 7.10 & 5.85& 7.1& 12.57\\
422754 & 26& 0.269& 4/7 & 3.28& 2.84 & 1.03 & 0.085 & 7.93 & 3.74 & 91.1  & 0.69 & 7.62 & 7.37& 6.7& 12.53\\
428177 & 31& 0.29& 5/9 & 5.16 & 2.90 & 1.28 & 0.079 & 5.28 & 2.98 & 27.2  & 0.50 & 9.02 & 7.65& 6.5& 12.39 \\
456326 & 20 & 0.25 & 4/5 & 3.75 & 1.21 & 0.72 & 0.101 & 9.22 & 2.53 & 41.37 & 0.64 & 5.34 & 4.2 & 6.3 & 12.32 \\
461785 & 24 & 0.25 & 4/6 & 3.90 & 1.15 & 0.70 & 0.092 & 6.75 & 3.05 &  82.97& 0.77 & 7.64 & 6.38 & 9.2 & 12.46 \\
\hline
\hline
473329 & 16 & 0.313 & 3/5 & 3.87 & 1.31 & 1.31 & 0.112 & 8.74 & 4.03 & 105.10 & 0.63 & 8.61 & 6.40 & 6 & 12.27\\
475619 & 15 & 0.467 & 5/7 & 2.61& 2.62 &2.62 & 0.116 & 4.20 & 1.81 & -1.0  & 0.76 & 4.64 & 3.96&  5.25& 12.41\\
479290 & 18 & 0.389 & 4/7 & 3.63& 4.25 & 2.11 & 0.104 & 8.07 & 4.65 & 18.1  & 0.64 & 8.50 & 6.37&  7.8& 12.30\\
482155 & 18 & 0.333 & 3/6 & 1.34& 1.04 & 1.58 & 0.116 & 4.20 & 2.19 & 21.8  & 0.69 & 2.51 & 2.03&  8.4& 12.35\\
494011 & 13 & 0.385 & 3/5 & 5.54& 2.54 &1.90 & 0.121 & 6.52 & 2.68 & 29.8  & 0.79 & 7.56 & 5.70&  7.4& 12.19\\
498522 & 16 & 0.313 & 3/5 & 4.02& 1.86 &1.31 & 0.112 & 4.81 & 2.53 & 13.5  & 0.65 & 5.18 & 5.09&  7.2& 12.28\\
501725 & 13 & 0.462 & 3/6 & 4.82 & 1.90 & 2.54 & 0.121 & 5.36 & 2.46 & 8.94 & 0.70 & 4.72 & 4.36 & 5.8 & 12.23 \\
503437 & 16 & 0.313 & 3/5 & 6.12& 1.74 &1.31 & 0.112 & 4.03 & 2.68 & 23.2 & 0.77 & 7.87 & 6.01& 7.3& 12.18\\
505586 & 19 & 0.368 & 4/7 & 3.32& 2.41 &1.90 & 0.103 & 4.29 & 1.69 & 11.8 & 0.80 & 3.64 & 3.14& 9& 12.15\\
506720 & 16 & 0.313 & 3/5 & 5.25& 2.42 &1.31 & 0.112 & 10.18 & 5.11 & 72.0 & 0.52 & 11.45 & 10.48& 7& 12.22\\
514829 & 12 & 0.417    & 4/5 & 4.01& 2.26 & 2.26 & 0.121 & 6.69 & 3.74 & 91.0 &0.73 & 6.67 & 4.0 & 5.7& 12.17\\
515695 & 14 & 0.357    & 4/5 & 3.69& 2.30 &1.69 & 0.118 & 6.35 & 2.14 & 61.3 &0.62 & 7.06 & 5.06& 5.6& 12.16\\
517271 & 15 & 0.333    & 3/5 & 4.79& 2.05 &1.47& 0.116 & 5.72 & 3.44 & 5.0 & 0.66 & 5.49 & 5.09& 5.5& 12.15\\
519311 & 14 & 0.428 & 4/6 & 5.74& 3.51 & 2.30 & 0.118 & 5.50 & 2.24 & -31.9 & 0.84 & 6.18 & 5.89& 7.5& 12.14\\
523889  & 16 & 0.375 & 5/6 & 3.99& 1.86 &1.86& 0.112 & 5.65 & 3.44 & -12.8  & 0.51  & 6.90 & 6.88& 5& 12.07\\
529365  & 11 & 0.456 & 3/5 & 3.96& 2.48 &2.48& 0.127 & 4.20 & 2.98 & -11.9  & 0.71  & 4.38 & 4.24& 4.5& 12.11\\
530330 & 13 & 0.385 & 3/5 & 3.72& 2.54 & 1.90& 0.121 & 8.43 & 2.83 & 38.9  &0.62  & 5.26 & 3.85& 4& 12.15\\
531910 & 11 & 0.545 & 4/6 & 5.81& 3.92 &3.20 & 0.127 & 7.06 & 3.59 & 122.7  & 0.63  & 7.77 & 5.87& 7.3& 12.05\\
535774 & 14 & 0.357  & 3/5 & 1.27& 2.30 &1.69& 0.118 & 6.35 & 2.14 & 23.5  & 0.82  & 4.63 & 4.35& 6.7& 12.0\\
543114 & 10 & 0.5    & 3/5 & 5.22&  2.00 &2.76& 0.132 & 4.03 & 1.46 & -23.0 &0.63 & 5.90 & 3.79& 4.5& 12.05\\
544408 & 11 & 0.456     & 3/5 & 5.61& 1.76 &2.48& 0.127 & 6.99 & 1.94 & -28.1  & 0.76  & 5.25 & 4.12& 7.2& 11.98\\
547293 & 13 & 0.615    & 4/8 & 3.24& 3.81 &3.81& 0.121 & 4.97 & 1.94 & 52.7 & 0.71 & 5.06 & 4.03& 6.1& 11.96\\
550149 & 9  & 0.667    & 4/6 & 3.04& 3.97 &3.97& 0.136 & 6.67 & 3.36 & 121.1  & 0.65 & 6.95 & 6.52& 5.3& 11.98\\

\hline
\end{tabular}
\label{ch4:tab:DataTable1}
\end{table*}


\vspace{-0.5cm}
\section{KPPs as skeletons of positional planes}
\label{ch4:sec:KPP_PosSpace}

\begin{figure*}
\centering
\includegraphics[width=0.48\linewidth]{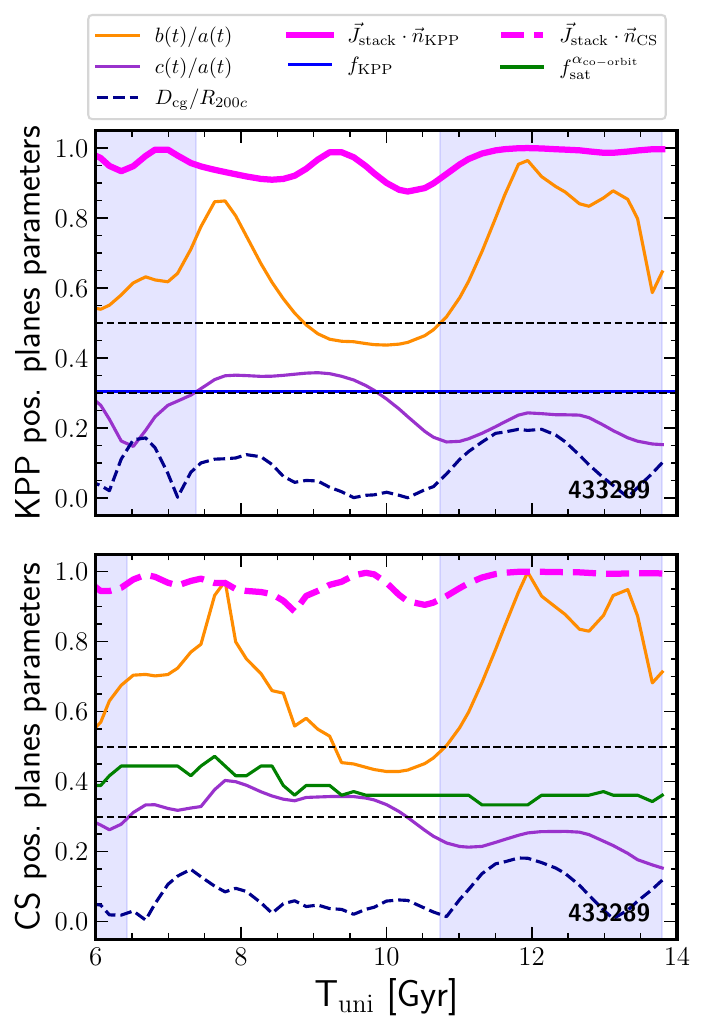}
\includegraphics[width=0.48\linewidth]{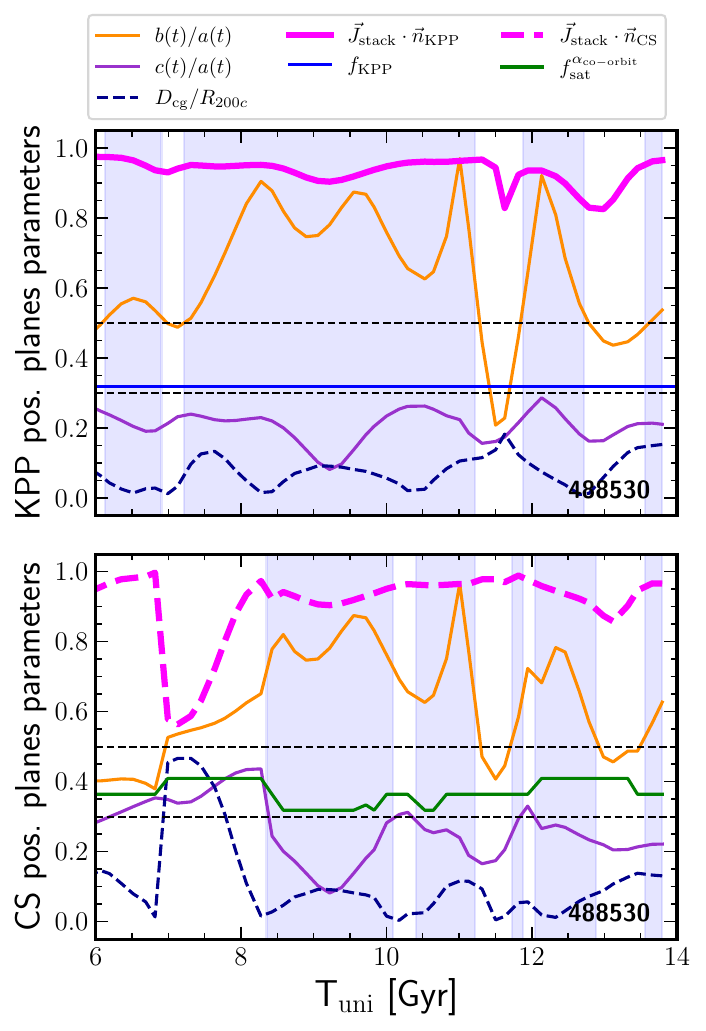}
\caption{Time evolution of the spatial planar structures formed by KPP satellites (upper panels) and CS sets (lower panels) in two example KPP-HS systems. Different lines represent different plane characteristics as described by the TOI analysis; i.e., $c/a$, $b/a$, $D_{\rm cg}/R_{\rm 200c}$, see legends. The fraction of satellites in KPPs ($f_{\rm KPP}$) and in CS sets ($f_{\rm sat}^{\alpha_{\rm co-orbit}}(t)$) are marked as blue and green lines, respectively. In addition, we show the cosine of the angle formed by the $\vec{J}_{\rm stack}$ vector and the normal vector to the KPP ($\vec{n}_{\rm KPP}$) and CS ($\vec{n}_{\rm CS}$) positional plane (magenta thick continuous and dashed lines, respectively). Results are shown for the period since T$_{\rm uni}\simeq6$ Gyr to $z=0$. Shaded vertical bands highlight those periods when the positional planes show thin and oblate morphologies (i.e., $c/a\leq0.3$ and $b/a\geq0.5$, see horizontal black, short-dashed lines).}
\label{ch4:fig:kineplane-ori}
\end{figure*}

In this section we assess the quality of KPPs as planar configurations in position space. 
To this end, we employ the standard Tensor of Inertia method \citep[TOI; see][]{Cramer}. An ellipsoid is fitted to the distribution of positions of satellites, either for KPP and CS sets, resulting in a planar configuration described by: their intermediate-to-long and short-to-long axis ratios ($b/a$ and $c/a$), their distance to the central galaxy c.o.m. ($D_{\rm cg}$), and their normal vectors ($\vec{n}_{\rm KPP}$ and $\vec{n}_{\rm CS}$).
%

%
%
%

We consider KPPs (or CS sets) in position space to be thin and  oblate if their axis ratios show values $c/a\leq0.3$ and $b/a\geq0.5$. Furthermore, their centroid should be located at close distances to the host center, as evidenced by low values of $D_{\rm cg}$. Moreover, their normal vectors $\vec{n}_{\rm KPP}$ (or $\vec{n}_{\rm CS}$) should align with the  $\vec{J}_{\rm stack}$ axis  of the corresponding KPP.


In the upper panels of Fig.~\ref{ch4:fig:kineplane-ori} we show some relevant results from the standard TOI analysis applied to the KPP satellites of two KPP-HS systems listed in Tab.~\ref{ch4:tab:DataTable1}.
Specifically, we plot the time evolution (from T$_{\rm uni}\simeq$ 6 Gyr onwards) of $c/a$, $b/a$ and $D_{\rm cg}/R_{\rm 200c}$. 
We also plot the cosine of the angle formed by the axis of maximum co-orbitation, $\vec{J}_{\rm stack}$, and $\vec{n}_{\rm KPP}$. Additionally, we plot the fraction $f_{\rm KPP}$ of satellites forming the KPP as a blue horizontal line. 
In the second row panels of Fig.~\ref{ch4:fig:kineplane-ori}, we present the same parameters for the CS sets of the same KPP-HS systems, alongside the fraction $f_{\rm sat}^{\alpha_{\rm co-orbit}}(t)$, represented as a green line.
%

For KPP satellites, Fig.~\ref{ch4:fig:kineplane-ori} illustrates the variation of plane spatial characteristics over time, revealing oscillations to different extents.
In the P24 KPP-HS systems, these oscillations show relatively long  periods, alternating  time intervals where the KPP positional 
plane remains thin and oblate, and others where $b(t)/a(t)$ is low  and $c(t)/a(t)$ increases. 
Indeed, some KPP-HS systems listed in Tab.~\ref{ch4:tab:DataTable1} show situations where $b(t)/a(t)\sim c(t)/a(t)$, i.e., the positional configuration becomes prolate/filamentary. This situation more frequently occurs in HS systems listed in the third block of Tab.~\ref{ch4:tab:DataTable1}.
In general, KPPs show a thin and oblate morphology for the bulk of the period analyzed; moreover, at specific times, their quality is exceptional. Those time intervals where positional planes are thin and oblate are  highlighted as shaded blue vertical bands in panels of Fig.~\ref{ch4:fig:kineplane-ori}.

\begin{figure*}
\centering
\includegraphics[height=6cm]{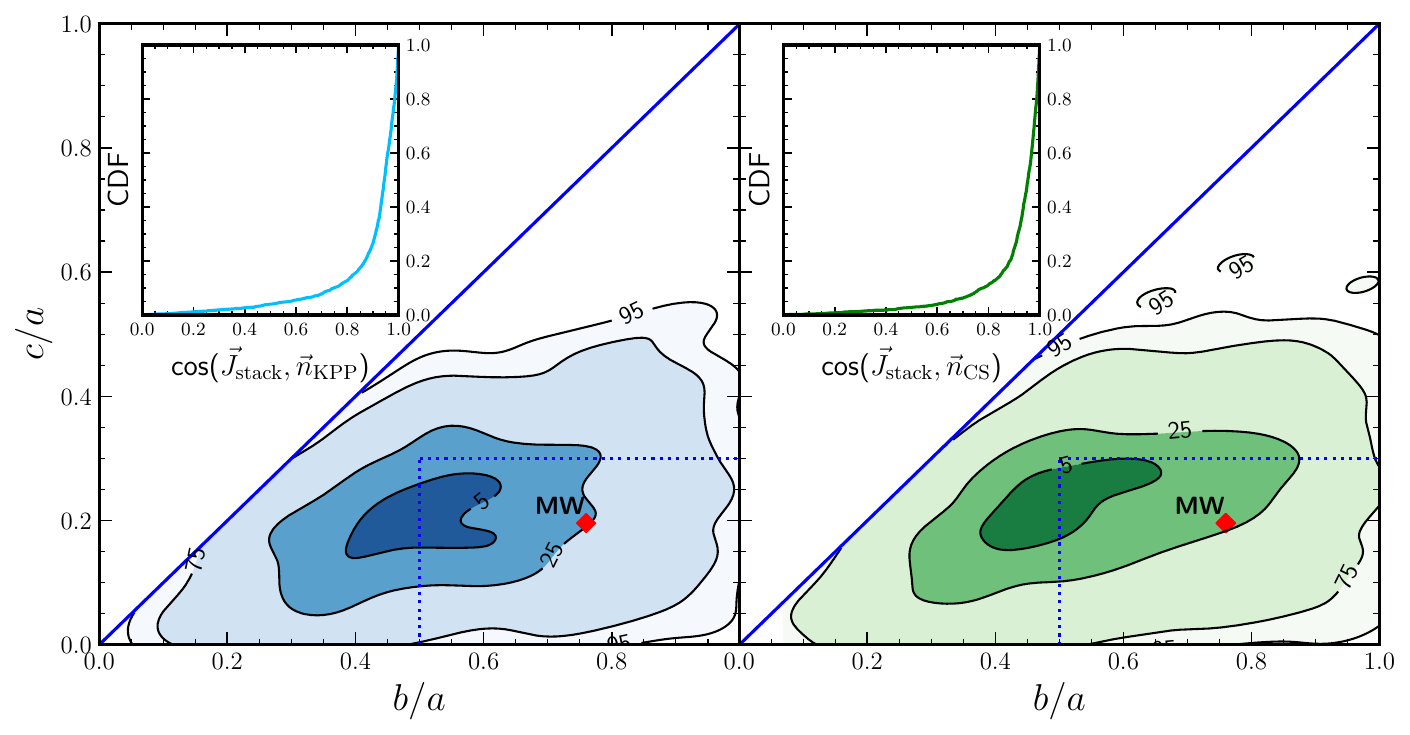}
\includegraphics[height=6cm]{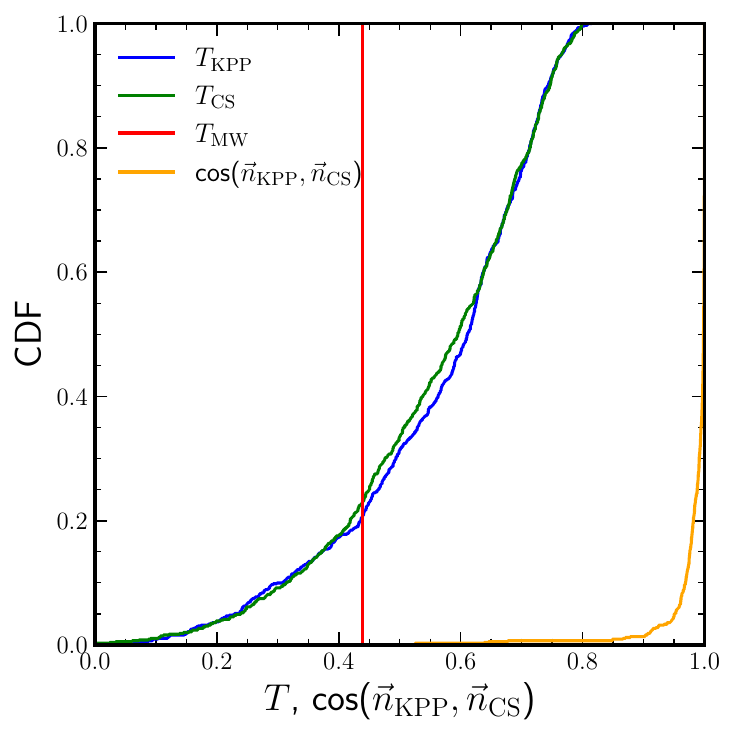}

\caption{\textit{Left and central panels:} Plots of the  $b/a$ and $c/a$ axes ratios for the 2254 positional planes in the statistical samples of KPP (left panel) and CS (central panel) sets. The iso-density contours enclose the 5\%, 25\%, 75\% and 95\% of the total number of fitted planes. 
As a visual guide, a rectangle delimited by blue-dotted lines highlights the parameter space corresponding to our definition of thin and oblate planes (i.e. $c/a\leq0.3$ and $b/a\geq 0.5$). 
The distributions of the cos($\vec{J}_{\rm stack}$,$\vec{n}_{\rm KPP}$) and cos($\vec{J}_{\rm stack}$,$\vec{n}_{\rm CS}$) values for the 2254 planes are presented as inset plots in their corresponding panels. The value of the axes ratios for the positional configuration  of satellites forming the kinematic MW plane is indicated by a red diamond, see Sec.~\ref{ch4:sec:Comparison_MW}.
\textit{Right panel:} CDF of the triaxiality values for thin and oblate planes (this is, located  simultaneously within the rectangle delimited by the blue-dotted lines, i.e., 752 out of 2254 planes) for both KPP satellites (blue line) and CS sets (green line). The orange line shows the alignment signal between $\vec{n}_{\rm KPP}$ and $\vec{n}_{\rm CS}$ vectors for these planes. The red line marks the $T$ value  of the MW kinematic plane.
}
\label{ch4:fig:Props_pos_planes}
\end{figure*}

As for the CS sets, the lower panels in Fig.~\ref{ch4:fig:kineplane-ori} show that their morphological evolution generally follows the same trends as their KPP satellite sets counterparts. This suggests that KPP satellites largely determine the morphology of positional planes fitted to CS sets. However, some differences between parameters may arise at specific time intervals.
As mentioned in Sec.~\ref{ch4:sec:Quant_SatCoorb}, CS sets contain on average 19\% more satellites than KPP satellite sets.

To quantify these trends in detail, for each KPP-HS system we collect the properties of the fitted planes of KPP and CS sets across all simulations outputs since T$_{\rm uni}\simeq 6$ Gyr,  totaling a sample of 2254 planes in each case (i.e., the statistical samples of KPP  and CS sets\footnote{The statistical sample of KPP satellites comes from the statistical sample of CS sets but considering KPP satellites instead of CS sets.}, respectively). 
First, we consider the two-variable distribution of $b/a$ and $c/a$ axis ratios.
In the left and central panels of Fig.~\ref{ch4:fig:Props_pos_planes} we present the probability density maps for the joint distributions of these axis ratios.  For each distribution, iso-density contours enclose 5\%, 25\%, 75\% and 95\% of the total number of fitted planes, as indicated by the labels on the iso-curves.

We see that the iso-curves exhibit similar patterns.  
In order to quantify potential differences between the distributions of $(b/a,c/a)$ values for KPP and CS sets, we compute the triaxility parameter $T=(1-(b/a)^2) / (1-(c/a)^2)$ \citep[][]{Franx:1991}. 
The right panel in Fig.~\ref{ch4:fig:Props_pos_planes} shows the CDFs of the $T$ parameters when, for a given KPP-HS at a specific time, both KPP and CS sets are thin and oblate 
(i.e., time intervals during which the blue shaded bands for both sets overlap). This selection leaves with 752 planes that, for both KPP and CS sets, locate within the rectangle defined by the blue-dotted lines in the left and middle panels in Fig.~\ref{ch4:fig:Props_pos_planes}. The comparison between $T$ results shows that their values are almost indistinguishable. Indeed, a Kolmogorov-Smirnoff (KS) test yields a p-value = 0.32, meaning that the null hypothesis cannot be rejected, confirming that KPPs predominantly shape the positional planes of CS sets in time periods in which they are thin and oblate.

Regarding the alignment between $\vec{J}_{\rm stack}$ and  $\vec{n}_{\rm KPP}$ vectors in the two examples shown in Fig.~\ref{ch4:fig:kineplane-ori}, we find that these two vectors are generally well aligned throughout the analyzed period (see magenta lines).
A similar trend is observed for the alignment between $\vec{J}_{\rm stack}$ and $\vec{n}_{\rm CS}$, especially in the time intervals when the planes are thin and oblate, marked by blue shade bands.

Let us now look at the statistics of these alignments. 
In the right panel of Fig.~\ref{ch4:fig:Props_pos_planes},  we display the CDF for the cos($\vec{n}_{\rm KPP}$, $\vec{n}_{\rm CS}$) values (orange line) for the thin and oblate planes set mentioned above (i.e., those encompassed within the rectangles in Fig.~\ref{ch4:fig:Props_pos_planes}). Only $\sim0.5\%$ of the 752 planes left are misaligned (i.e., cos($\vec{n}_{\rm KPP}$, $\vec{n}_{\rm CS}$) $<$ 0.8), an indication that KPP orientation determines CS orientation. 
Additionally, in the left and middle panels of Fig.~\ref{ch4:fig:Props_pos_planes} we plot as insets the CDFs of the alignment values between $\vec{J}_{\rm stack}$ and the normal vectors to the positional planes, $\vec{n}_{\rm KPP}$ and $\vec{n}_{\rm CS}$.
We see in each case that the fraction of non-aligned kinematic and positional planes is low, proving that the result shown in Fig.~\ref{ch4:fig:kineplane-ori} for individual KPP-HS systems overall holds statistically for the entire population of KPP-HS systems. 
Thereby, we confirm the high coherence of directions of the $\vec{J}_{\rm stack}$, $\vec{n}_{\rm KPP}$ and $\vec{n}_{\rm CS}$ vectors during long time intervals. 
This is a very relevant result, and it indicates that KPPs not only shape CS sets, but they also determine to a great extent the directions of their normals $\vec{n}_{\rm CS}$ as positional planes. In summary, KPPs act as a kind of skeleton for CS sets.


Regarding the positional plane distance to the central galaxy, $D_{\rm cg}(t)/R_{\rm 200c}$ has an oscillating behavior too, both for KPP and CS structures.  The latter show sub-fluctuations as well, presumably caused by non-persistent satellites going in and out the structure. During the slow halo mass-assembly phase, its minimum values are often $\sim$ 0, while maxima generally remain below $\sim$  20\%  of the virial radius. The 6 HS systems in blocks 1 and 2 of Tab.~\ref{ch4:tab:DataTable1} that exceed $\sim 0.2$ have $N_{\rm KPP} \leq 7$. Some reasons responsible for this relatively larger $D_{\rm cg}$ might be: i), late infall timescales, where some KPP satellites are still in their infall process, ii), HS systems are embedded in a highly dynamical active environment, including situations where the KPP temporarily adopts a filamentary/prolate  shape (i.e., $b \sim c$), and iii), situations where all KPP satellite members are, at the same time,  radially far from the host, akin to apocenter accumulation.
Notably, these factors lead to heavily lopsided satellite distributions, consistent with findings in Paper III, where we confirmed that relatively high $D_{\rm cg}/R_{\rm 200c}$ values coincided with asymmetric satellite distributions.

%

We finish this subsection by noting that had we carried out a search for planes of satellites in these KPP-HS systems  based only on positional information of satellites, we would have yielded more populated positional planes (i.e., more satellites in the positional plane  for equivalent $c/a$ values; see e.g. Paper II). These additional satellites are interlopers or  transient members of the plane without their orbital poles co-orbiting at that time, as mentioned in Sec.~\ref{ch4:sec:intro}. 


\begin{figure*}
\centering
\includegraphics[height=5.3cm]{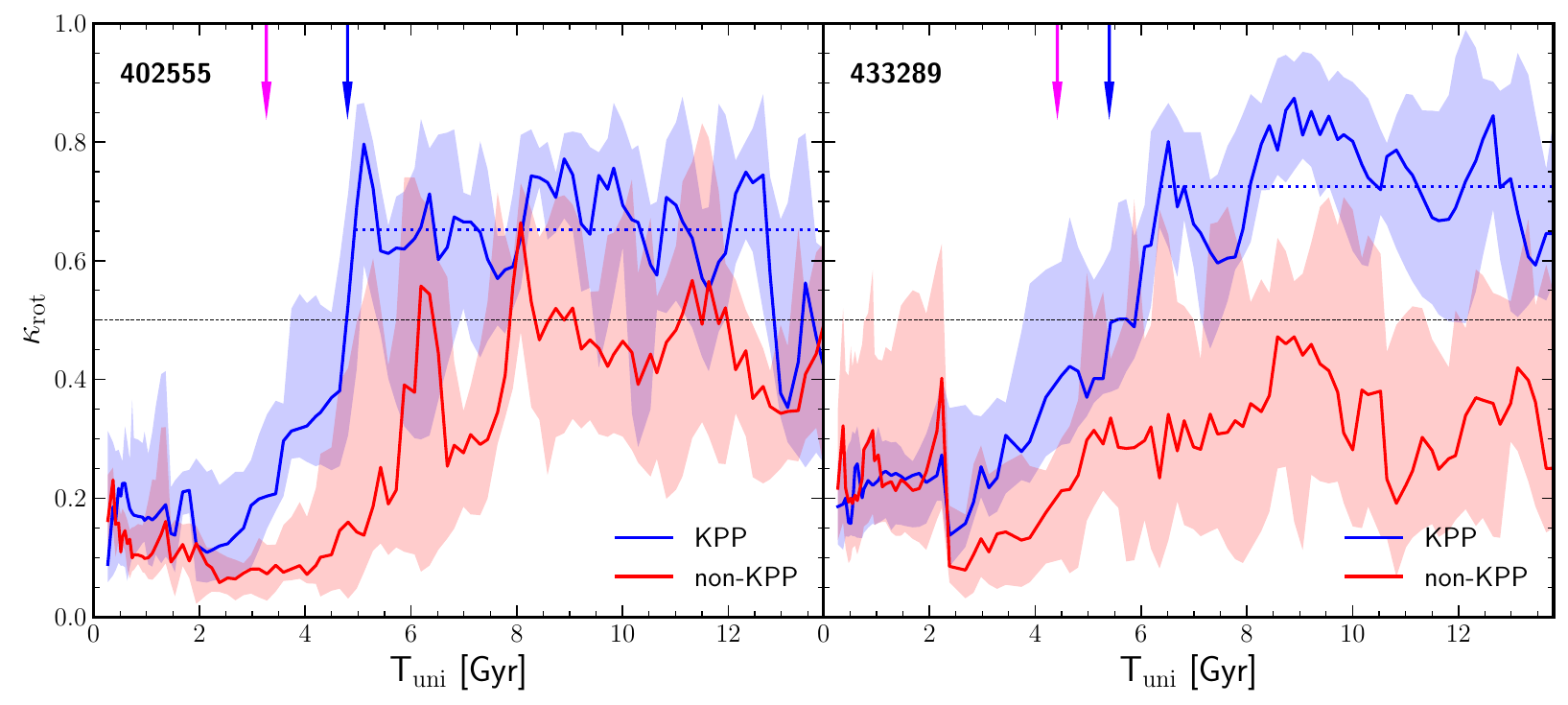}
\includegraphics[height=5.3cm]{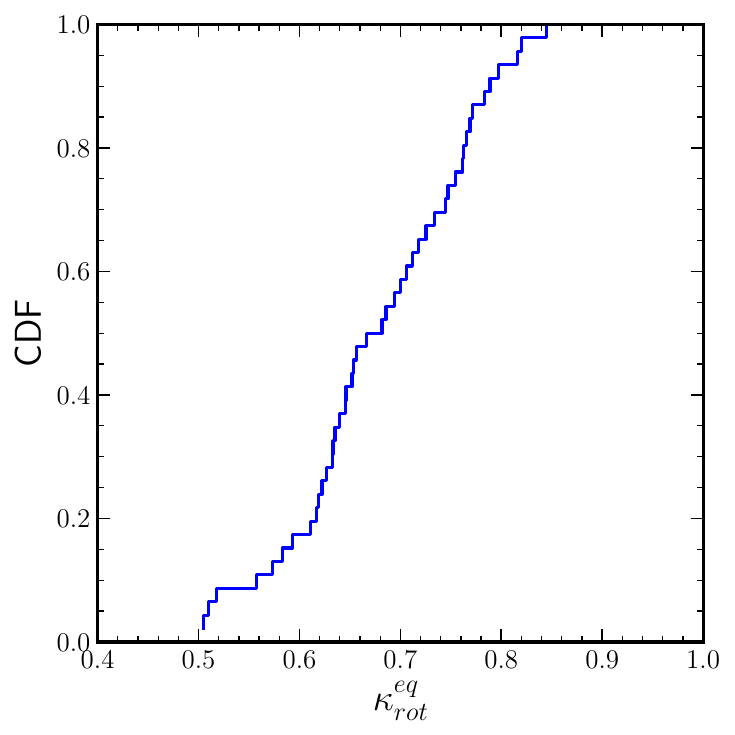}
\caption{\textit{Left and central panels}: The kinematic morphological parameter $\kappa_{\rm rot}(t)$ curves for satellites belonging to two KPP-HS systems from  Tab.~\ref{ch4:tab:DataTable1}. Blue (red) curves display the median $\kappa_{\rm rot}(t)$ values for  KPP (non-KPP) satellites; shaded areas represent the corresponding $25-75$th percentiles. 
 Magenta and blue arrows mark the $T_{\rm cluster}^{\rm Jstack}$ and $T_{\rm no-fast}$ timescales of each halo, respectively.
Blue horizontal lines give the $\kappa_{\rm rot}^{\rm eq}$ values in each case. Grey horizontal lines mark the  $\kappa_{\rm rot} = 0.5$ value,
considered as the limit between rotation-dominated and dispersion-dominated systems. 
\textit{Right panel}: Cumulative distribution of the $\kappa_{\rm rot}^{\rm eq}$ values for  all KPP-HS systems listed in Tab.~\ref{ch4:tab:DataTable1}.
} 
\label{ch4:fig:Krot-median-satellites}
\end{figure*}

\section{KPPs in Velocity Space}
\label{ch4:sec:krot}

In this section we study the motion of  satellites in velocity space making use of the kinematic morphological parameter, $\kappa_{\rm rot}$.
Originally defined by \citet{Sales:2012} to morphologically classify galaxies, 
in Paper IV we extended its definition to satellite systems, analyzing it relative to the $\vec{J}_{\rm stack}$ axis. As such, for a given $i-$th satellite in a given HS system, the energy ratio in ordered motion (i.e., in-plane and circularized) at an Universe age T$_{\rm uni}=t$ can be written as  $\kappa_{\mathrm{rot},i}(t)=(v_{i,\phi} / v_i )^2(t)$, where $v_{i,\phi}$ represents the satellite tangential velocity (in cylindrical coordinates) relative to the system's center of velocity in the plane normal to the system's $\vec{J}_{\rm stack}$, while $v_i$  is the module of its velocity. In this way, the $\kappa_{\rm rot}$ parameter (the median of the $\kappa_{\mathrm{rot},i}$ values) serves as a kinematic indicator of the morphology of a  satellite set: dispersion dominated ($\kappa_{\rm rot}$ < 0.5) or rotation dominated (also termed disky-like motion in this context,  $\kappa_{\rm rot}$ > 0.5). 
\footnote{It is worth noting that, contrary to the original definition of $\kappa_{\rm rot}$, in this analysis the mass of individual satellites do not play any dynamical role and hence mass is not used to weight the contribution of each satellite to $\kappa_{\rm rot}$.}

Results for two example KPP-HS systems are given in the left and middle panels of Fig.~\ref{ch4:fig:Krot-median-satellites}. These panels depict the time evolution of the $\kappa_{\rm rot}(t)$  curve for KPP satellites (blue line) and non-KPP (red line) satellites. Shaded bands show their corresponding $25-75$th percentiles.
Magenta and blue arrows mark the $T_{\rm cluster}^{\rm Jstack}$ and $T_{\rm no-fast}$ timescales of each KPP-HS system, respectively.

A first characteristic result from these figures is that KPP satellites present low values of  $\kappa_{\rm rot}(t)$ at early times, during the fast phase of halo mass assembly, prior to orbital pole clustering timescales (see arrows). Then, as time elapses, the $\kappa_{\rm rot}(t)$ curve increases, reaching values up to 0.5, implying disky-like motion from that moment onwards.
These values then stabilize to roughly constant median values, allowing us to define the $\kappa_{\rm rot}^{\rm eq}$ parameter for KPP satellites, computed as the mean value of the $\kappa_{\rm rot}(t)$ values from $T_{\rm Krot}^{\rm eq}$ (see definition below) onwards. The $\kappa_{\rm rot}^{\rm eq}$ value is indicated by the horizontal blue dotted line in the left  and middle panels of Fig.~\ref{ch4:fig:Krot-median-satellites}. Tab.~\ref{ch4:tab:DataTable1} provides the $\kappa_{\rm rot}^{\rm eq}$ values of all KPP-HS systems. 
As seen in these examples, the $\kappa_{\rm rot}(t)$ function is time-fluctuating, primarily due to low-number statistics coupled with  peaks in individual satellite curves at their apocenter and pericenter positions. 


%
The right panel of Fig.~\ref{ch4:fig:Krot-median-satellites} shows the CDF of $\kappa_{\rm rot}^{\rm eq}$ found for all KPP-HS systems. KPP satellites generally exhibit high values of this parameter, indicating that these satellite populations have disk-like behaviors from a kinematical perspective.
Conversely, $\kappa_{\rm rot}(t)$ values for non-KPP satellites remain below  0.5 at any time, with only a few occasional peaks above this value. 
This indicates a clear segregation in terms of $\kappa_{\rm rot}$. 

We define a timescale for the onset of the ordered disky motion in a specified KPP-HS system, $T_{\rm Krot}^{0.5}$, as the Universe age when $\kappa_{\rm rot}(t)$ = 0.5 for the first time. 
A second timescale, $T_{\rm Krot}^{\rm eq}$, is given by the Universe age when the $\kappa_{\rm rot}(t)$ curve for KPP satellites stops increasing, except for fluctuations, meaning that the `disky' KPP satellite configuration is established at its maximum circularity value. In some cases, 
a fast setting in of the planar disky  configuration occurs, i.e., the average  slope of the $\kappa_{\rm rot}(t)$ curve 
before $T_{\rm Krot}^{0.5}$ is high
(see, e.g., the left panel of Fig.~\ref{ch4:fig:Krot-median-satellites}).
In other cases, it takes longer for KPP satellites to acquire their disky  configuration, as shown in the middle panel of Fig.~\ref{ch4:fig:Krot-median-satellites}. 
Values for both kinematic timescales, $T_{\rm Krot}^{0.5}$ and  $T_{\rm Krot}^{\rm eq}$, are listed in Tab.~\ref{ch4:tab:DataTable1}, and their CDFs are provided in Fig.~\ref{ch4:fig:CDF-timescales}, and Fig.~\ref{ch4:fig:Timescales_visual_distrib_ch4} as a visual guide. 
We see that $T_{\rm Krot}^{\rm eq} \sim T_{\rm no-fast}$, and that the former comes generally after $T_{\rm cluster}^{\rm Jstack}$ (i.e., the timescale for orbital pole clustering related to  satellite co-planar motion acquisition), suggesting that in-plane motion is stablished generally before the circularization of satellite orbits. More details about this point will be provided in the next section.

%


\section{Analyses of timescale relationships and statistics}
\label{ch4:sec:VariableRelationships}

\begin{figure*}
\centering
\includegraphics[width=0.8\linewidth]{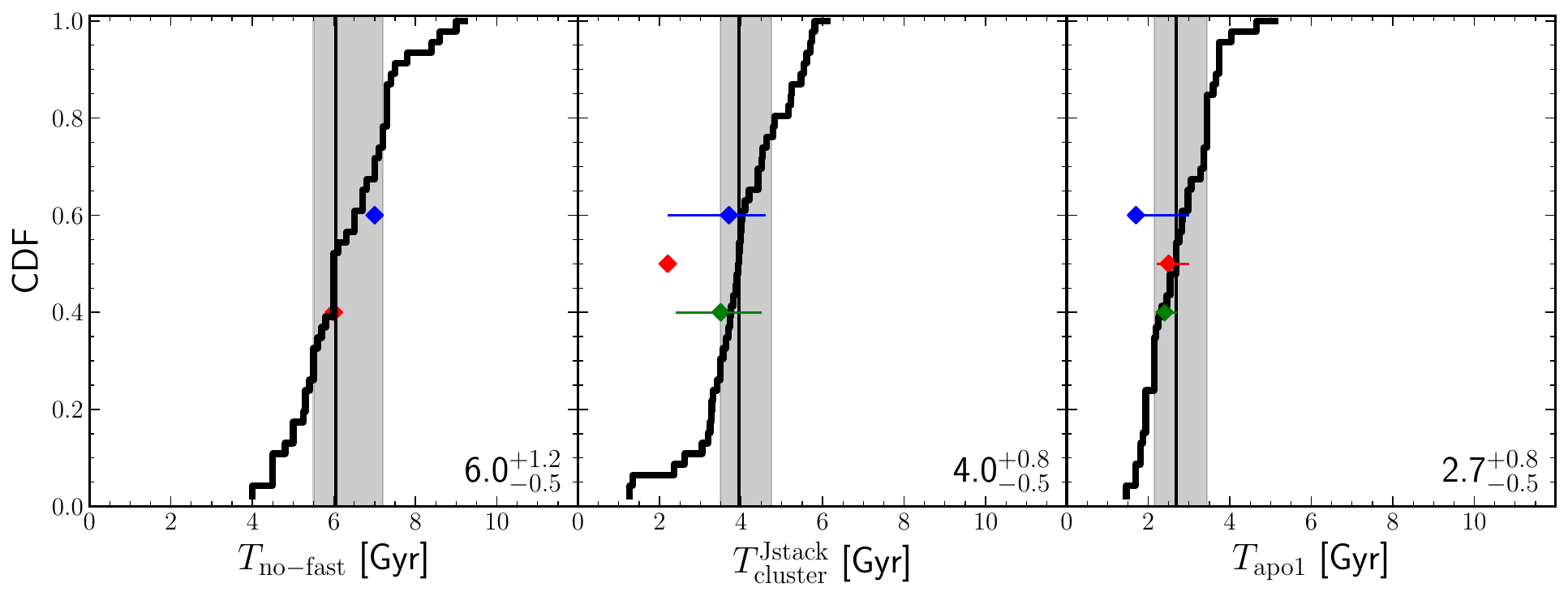}
\includegraphics[width=0.8\linewidth]{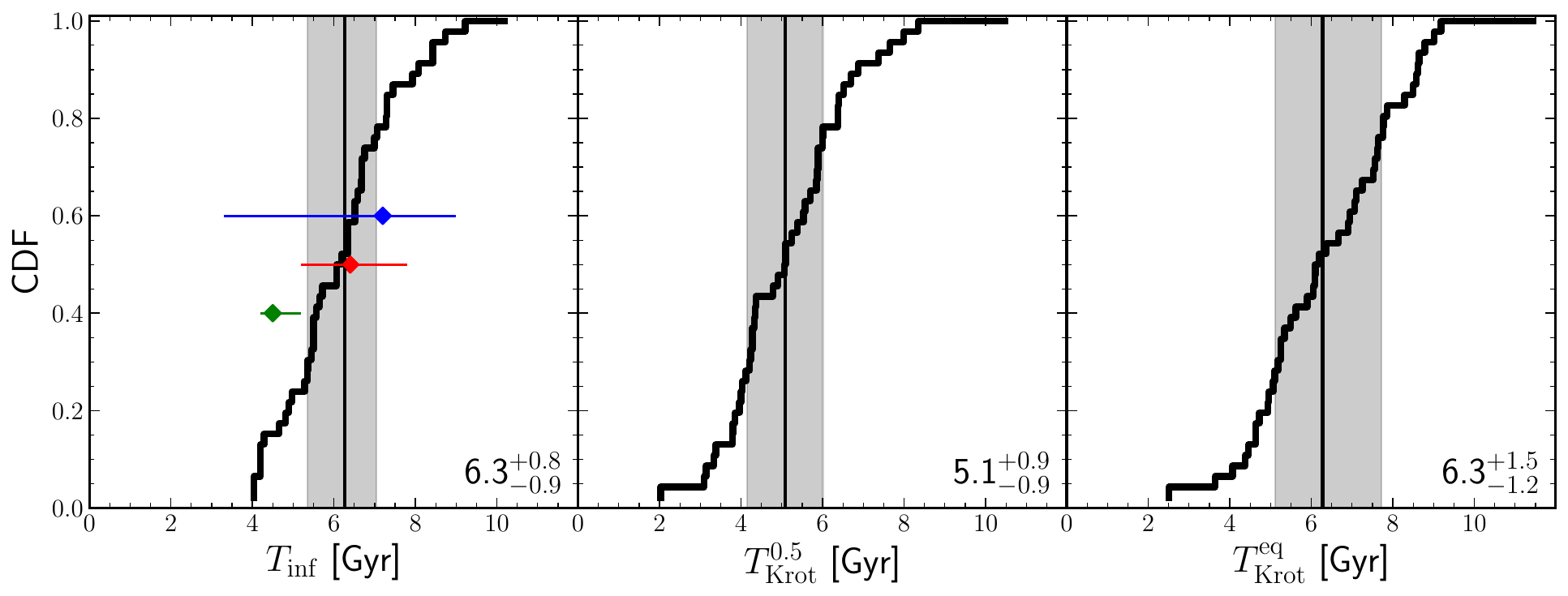}

\caption{CDFs of the different timescales analyzed in this paper. The median and $25-75$th percentiles for each of the analyzed distributions are depicted in the lower right corner of each plot, as well as visually depicted as thick lines and shaded vertical bands, respectively. 
See main text for the definitions of each timescale. 
Diamonds with errorbars show the corresponding results obtained with the zoom-in simulations studied in Papers III and IV. Specifically, one KPP was identified in the Aq-C$^{\alpha}$ simulation (shown in blue) and two KPPs were identified in the PDEVA-5004 simulation (red and green).
}

\label{ch4:fig:CDF-timescales}
\end{figure*}

Throughout this paper we have presented different timescales concerning the properties along evolution of HS systems and the formation of KPPs within some of them.
In this section, we analyze the possible relationships among these and other processes by comparing the timescales at which they occur. 

\begin{figure}
\centering
\includegraphics[width=0.9\linewidth]{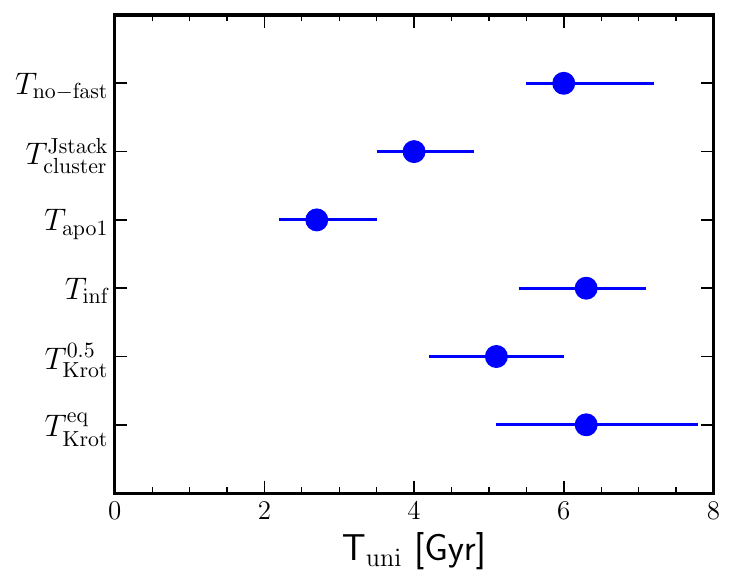}
\caption{Representation of the distributions of the different timescales analyzed in this paper. Median values and $25-75$th percentiles are represented by blue filled points with horizontal error bars.}
\label{ch4:fig:Timescales_visual_distrib_ch4}
\end{figure}

In Fig.~\ref{ch4:fig:CDF-timescales} we represent the CDFs of the analysed timescales. For each timescale, the median and $25-75$th percentiles are written on the bottom right corner of each panel, and also marked as black vertical lines and grey shaded regions, respectively. 
A visual guide of the distribution of these timescales is also shown in Fig.~\ref{ch4:fig:Timescales_visual_distrib_ch4}.
Additionally, we have applied a Spearman's test to all pairs of timescales, as an analytical tool to assess the strength of possible correlations.


Here we enumerate several of the most relevant results derived from these analyses:

\begin{enumerate}[label = {\roman*)}, wide, left=0pt, labelsep=0em,labelindent=0pt]
    \item According to Fig.~\ref{ch4:fig:Timescales_visual_distrib_ch4}, the segments representing $T_{\rm no-fast}$, $T_{\rm inf}$, and $T_{\rm Krot}^{\rm eq}$ show very close median values ($\sim6$ Gyr), while the segment representing $T_{\rm Krot}^{0.5}$, has a median value slightly lower (around 1 Gyr lower) than the other three timescales. The physics behind the former coincidences can be explained as follows:
    \vspace{-0.2cm}
    \begin{itemize}[labelsep=0.5em]
        \item The coincidence between $T_{\rm no-fast}$ and $T_{\rm Krot}^{\rm eq}$ indicates that, once the violent dynamical activity halts at $T_{\rm no-fast}$, the loss of mechanical energy by satellites is hindered, hence stabilizing their orbits at $T_{\rm Krot}^{\rm eq}$. This coincides, on average, with the moment when satellites enter into the host halo at $T_{\rm inf}$.
        \item The time gap between $T_{\rm Krot}^{0.5}$ and $T_{\rm Krot}^{\rm eq}$ varies from $\sim0$ to $\sim 1.5$ Gyrs, indicating that orbit circularization can be either rapid or gradual after KPP formation, as already mentioned.
        \item The dispersions of $T_{\rm no-fast}$, $T_{\rm inf}$, and $T_{\rm Krot}^{0.5}$ are similar and very narrow ($\sim0.85$ Gyr). In contrast, $T_{\rm Krot}^{\rm eq}$ shows a broader distribution ($\sim1.35$ Gyr), given the range of possible time gaps between $T_{\rm Krot}^{0.5}$ and $T_{\rm Krot}^{\rm eq}$ explained above.
    \end{itemize}
\vspace{-0.2cm}
    \item $T_{\rm apo1}$, $T_{\rm cluster}^{\rm Jstack}$ and $T_{\rm no-fast}$ timescale ranges in Fig.~\ref{ch4:fig:Timescales_visual_distrib_ch4} do not overlap, reflecting distinct evolutionary phases,
    namely satellite turn-around, KPP formation, and the end of the fast-phase of halo assembly, respectively.
 
    \item The ranges of  $T_{\rm cluster}^{\rm Jstack}$ and  $T_{\rm Krot}^{0.5}$ do overlap, indicating that, as KPPs form and satellites gain in-plane motion, satellites slowly start to circularize their orbits.
    
    \item The inequality $T_{\rm cluster}^{\rm Jstack} < T_{\rm no-fast}$ shows that orbital pole clustering typically occurs during the host’s fast mass assembly phase. This supports the idea that early pole clustering is not driven by halo processes. Likewise, $T_{\rm cluster}^{\rm Jstack} < T_{\rm inf}$ reinforces this interpretation. The fact that $T_{\rm Krot}^{\rm eq} > T_{\rm cluster}^{\rm Jstack}$ suggests that KPP formation precedes orbital circularization and stabilization in KPP satellites.

    \item Narrow timescale dispersions preclude strong correlations between them, with the exception of $T_{\rm Krot}^{0.5}$ with $T_{\rm Krot}^{\rm eq}$ (see point i) above).

\end{enumerate}

These results reinforce the idea, hinted in previous sections, of satellites
establishing their kinematic coherence at very early times, when the proto-satellites are but part of the galaxy-to-be evolving environment, moving at unison with it. 


\section{Discussion}
\label{ch4:sec:Discussion}

\subsection{Possible satellite mass effects}
\label{ch4:sec:Mass-Effects-KPPvsnonKPP}

Throughout this work, we have analyzed different parameters that distinguish KPP and non-KPP satellite populations within a given KPP-HS system. Satellite mass could also be a parameter determining a satellite's KPP or non-KPP membership. The interest of analyzing this point lies in the inherent limited mass resolution of simulations.

\begin{figure}
\centering
\includegraphics[width=0.9\linewidth]{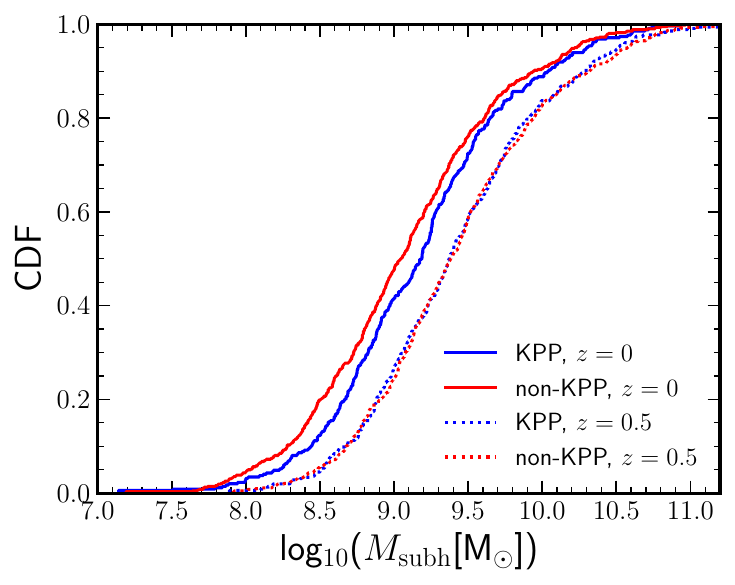} 
\caption{CDFs for the $M_{\rm subh}$ masses of satellite populations in the KPP-HS systems listed in Tab.~\ref{ch4:tab:DataTable1}. Blue (red) curves stand for KPP (non-KPP) satellite members.
Solid (dotted) lines stand for satellite mass measured at $z=0$ ($z=0.5)$.
}
\label{ch4:fig:CDF-SatMass-All}
\end{figure}

Fig.~\ref{ch4:fig:CDF-SatMass-All} shows the cumulative distributions of satellite $M_{\rm subh}$ mass at $z=0$ (solid lines) and at $z=0.5$ (T$_{\rm uni}\simeq 8.58$ Gyr, dotted lines) for all KPP and non-KPP satellite subsamples from all KPP-HS systems. 

Looking at the CDF mass curves of both populations at $z=0$, a KS test distinguishes KPP from non-KPP $z=0$ populations at a 91\% confidence level, with KPP satellites being slightly more massive. 
However, this distinction between populations vanishes when comparing the  mass curves at $z=0.5$, where both populations exhibit an identical distribution of $M_{\rm subh}$\footnote{The comparison at $z=0.5$ is particularly suited for this analysis as, by this time, the majority of satellites have been captured by their host halo, see $T_{\rm inf}$ parameter distribution in Fig.~\ref{ch4:fig:CDF-timescales}.}.
From this analysis, we observe that while both populations have suffered from mass loss in the time interval between $z=0.5-0$, non-KPP populations have lost a higher fraction of their satellite mass $M_{\rm subh}$. 

We conclude that, in the P24 HS system sample, satellite halo mass ($M_{\rm subh}$) at infall is not a parameter playing any role at determining whether or not this satellite belongs to a KPP structure or not,
a conclusion already reached in Paper III.
Indeed, by the time when satellites are captured by their host, both KPP and non-KPP satellites have indistinguishable mass distributions. 


\subsection{Effects of satellite and KPP selection criteria }
\label{ch4:sec:Disc_sel_criteria}

In order to test the robustness of our results, we explore how varying the selection criteria used to identify both KPPs and satellites affect the $f_{\rm KPP-HS}^{\rm P24}$ and $f_{\rm KPP-HS}^{\rm P24-eli}$ fractions. 
Specifically, we assess the influence of changing the minimum stellar mass cut for satellites, $M_{\rm \ast,cut}$, and the minimum number of satellites required to form a KPP, $N_{\rm KPP}^{\rm min}$.
A summary of our numerical results is shown in Tab.~\ref{ch4:tab:FrequenciesTable}.

\paragraph*{\boldsymbol{$M_{\rm \ast,cut}$}}: We have decreased the stellar mass cut criteria used to define satellites, to $M_{\rm \ast,cut} = 4.25\times10^5$ M$_{\odot}$ (i.e. 5 stellar particles). This change introduces lower-mass satellites supplementing those satellites already considered in our analysis. After applying the $\vec{J}_{\rm stack}$ method to the expanded satellite populations, we find that 70 HS systems now host KPPs that fulfill the KPP selection criteria applied in this study ($N_{\rm KPP}\geq5$ and $f_{\rm KPP}\geq25\%$), increasing $f_{\rm KPP-HS}^{\rm P24}$ from 24.2\% to 36.8\%. 
When changing $M_{\rm \ast,cut}$, the number of \textit{eligible} HS systems also changes. Indeed, with this value of $M_{\rm \ast,cut}$, we are able to identify KPPs in HS systems with an $N_{\rm sat}^{\rm min}$ of $\geq7$ satellites.
This increases the number of eligible HS systems from 123 to 168, giving a value of $f_{\rm KPP-HS}^{\rm P24-eli}=41.6\%$. This value is similar to  the one obtained with our fiducial selection criteria, $f_{\rm KPP-HS}^{\rm P24-eli}=37.4\%$. These results suggest that increasing the mass resolution of simulations (while keeping the same KPP selection criteria) would only significantly increase $f_{\rm KPP-HS}^{\rm P24}$ but not $f_{\rm KPP-HS}^{\rm P24-eli}$.

\paragraph*{\boldsymbol{$N_{\rm KPP}^{\rm min}$}}: We have also analyzed the impact that 
changing our condition for the minimum number of satellites forming a KPP ($N_{\rm KPP}^{\rm min}$) may have
on our calculated frequencies. It turns out that, just by changing  $N_{\rm KPP}^{\rm min}$ from 5 to 6, $f_{\rm KPP-HS}^{\rm P24}$ significantly decreases from 24\% to $\sim 17\%$, while the frequency relative to the eligible HS systems changes from $f_{\rm KPP-HS}^{\rm P24-eli}=37.4\%$ to 26\%. It is worth noting that, in this case, the number of eligible HS systems remains constant.

\begin{table}
\caption{KPP frequencies relative to the total P24 HS systems ($f_{\rm KPP-HS}^{\rm P24}$) and to the \textit{eligible} HS systems ($f_{\rm KPP-HS}^{\rm P24-eli}$) for different values of the minimum stellar mass cut of satellites ($M_{\rm \ast,cut}$) and of the minimum number of satellites to consider a KPP ($N_{\rm KPP}^{\rm min}$). Parameters are changed one at a time relative to their fiducial value in this work ($M_{\rm \ast,cut}\geq8.5\times10^5$M$_{\odot}$, $N_{\rm KPP}^{\rm min}=5$). According to our selection criteria, eligible HS systems happen to have $N_{\rm sat}\geq9$.
}
\scriptsize
\hspace{0cm}
%
\vspace{0cm}
\begin{tabular}{|l | c  | c | c |}
\toprule
\multicolumn{1}{|c|}{Frequencies} & \multicolumn{1}{|c|}{This work} & \multicolumn{1}{c|}{$M_{\rm \ast , cut}\geq4.25\times10^5$M$_{\odot}$} 
& \multicolumn{1}{c|}{$N_{\rm KPP}^{\rm min}=6$}  \\
\hline
\multicolumn{1}{|c|}{$f_{\rm KPP-HS}^{\rm P24}$} &  \multicolumn{1}{c|}{24.2\% (46/190)} & \multicolumn{1}{c|}{36.8\% (70/190)} 
& \multicolumn{1}{c|}{16.8\% (32/190)} \\
\hline
\multicolumn{1}{|c|}{$f_{\rm KPP-HS}^{\rm P24-eli}$} & \multicolumn{1}{c|}{37.4\% (46/123)}   & \multicolumn{1}{c|}{41.6\% (70/168)}  
& \multicolumn{1}{c|}{26\% (32/123)}  \\
\bottomrule
\end{tabular}
\label{ch4:tab:FrequenciesTable}
\end{table}

These findings indicate that while the frequency of KPPs decreases as we demand more populated ones,  increasing the mass resolution of simulations could improve the accuracy of these frequencies by including more satellites per system (hence more eligible HS systems).


\subsection{Comparison to observational data}
\label{ch4:sec:Comparison_MW}

In this section we briefly compare the available data for MW satellites with our findings for CS sets around MW/M31-like hosts in TNG50,
see Figs.~\ref{ch4:fig:unocos} (c), \ref{ch4:fig:unocos} (d) and \ref{ch4:fig:Props_pos_planes}\footnote{The comparison has to be made with results of CS sets, as we ignore which of the MW's satellites are temporarily or persistently co-orbiting in a kinematically-coherent manner.}. 
Thanks to the \textit{Gaia} mission \citep{GaiaEDR3}, there are currently estimates for the proper motions of 50 MW satellites \citep[see][and references therein]{Battaglia22}.
With this data, we can compute satellite orbital pole directions, the axis of maximum co-orbitation, and determine $f_{\rm sat}^{\rm exc}(z=0)$ for the MW ($f_{\rm sat,MW}^{\rm exc}$).
An important aspect here is that the significant uncertainties in MW satellite proper motion data must be accounted for in our computations\footnote{Specifically, we use the sample of 50 satellites listed in table 2 of \citet{Taibi24}, computing orbital pole directions based on their Galactocentric 3D positions and velocities. To account for these data uncertainties, we calculate the maximum satellite co-orbitation axis for 5000 randomized orbital poles of these satellites, considering these uncertainties as Gaussian distributions. This approach yields a distribution of $f_{\rm sat,MW}$ values for each randomized configuration, from which we derive their mean and standard deviation. We will refer to the `most likely' values as the results computed from the measured central phase-space values of each MW satellite.}.
We find that 21 out of 50 satellites co-orbit following our methodology.

Some words of caution regarding the comparison between data and TNG50 are in order, however, as the actual MW system differs from the simulated host systems analyzed in this work in several respects. For example, the MW's specific formation history may have influenced its satellite population and distribution. Moreover, the binary Local Group environment with M31, and the recent infall of the LMC could also have  played a role.
Additionally, observational uncertainties further complicate direct comparisons;
for instance, the incompleteness in the census of MW satellites \citep[see discussion in][]{SantosSantos2024_anisotropic}.
%
%
Also, while the range of $N_{\rm CS}$ values (i.e., the number of satellites in a given CS set) in the statistical sample of CS sets reaches up to 26, and the dependence on $N_{\rm CS}$ and/or $N_{\rm sat}$ of the different parameters analyzed is weak (see Appendix~\ref{ch4:appendixB} and Tab.~\ref{ch4:tab:Correlations_Ncs}), only a few percent of planes from the statistical sample present an $N_{\rm CS}$ similar to 21
\citep[the number of co-orbiting MW satellites obtained from data in][]{Taibi24}.
Given these circumstances, we do not aim for quantitative agreement with MW data (as noted in Sec.~\ref{ch4:sec:intro}), but rather to identify qualitative similarities that may indicate the presence of common underlying physics.


In  Fig.~\ref{ch4:fig:unocos} (c) we plot the distributions of $f_{\rm KPP}^{\rm exc}$ and $f_{\rm sat}^{\rm exc}(t)$ for each KPP-HS system from the P24 sample, see Eq.~\ref{ch4:eq:excess-def} for their definition, in units of the dispersion $\sigma$ of the respective randomized distribution (see Tab.~\ref{ch4:tab:DataTable1}). Using these units, the $N_{\rm sat}$ dependence of the aforementioned quantities is mitigated (Spearman correlation coefficient $\rho=0.184$), allowing results for HS systems with different $N_{\rm sat}$ to be directly compared with the MW results.
The red diamond symbol in Fig.~\ref{ch4:fig:unocos} (c) illustrates the result: $f_{\rm sat,MW}^{\rm exc} = (4.06 \pm 0.99) \sigma$. This is encompassed within $f_{\rm sat}^{\rm exc}(t)$ global distribution values (black histogram).

The fraction of co-rotating satellites forming the kinematical plane in the MW is
$f_{\rm co-rot}^{\rm MW}=0.81$ (red dashed vertical line  in Fig.~\ref{ch4:fig:unocos} (d)). 
It is interesting to note that about a 11.6\%  of the KPP-HS systems studied here have $f_{\rm co-rot}$ for their CS sets higher than  $\simeq$ 0.8, so that the MW is not an outlier regarding co-rotation rates in the P24 sample.

Another interesting comparison between our results and the MW data can be read in Fig.~\ref{ch4:fig:Props_pos_planes}.
The morphological properties of the spatial plane formed by the MW's 21 kinematically-coherent satellites are depicted in the  central panel of Fig.~\ref{ch4:fig:Props_pos_planes} as a red diamond (error bars are too small to be seen). 
We find that the morphology of the MW plane is not uncommon among our CS planes. As for the triaxiality parameter, we estimate $T_{\rm MW}=0.43$. This means that 23\% of thin and oblate CS planes are more oblate than the MW plane. 

Overall, our results are in broad consistency with those for the MW. 
This consistency could indicate that our Galaxy's satellite system may be compatible with $\Lambda$CDM predictions, though -- given the paucity of the P24 results close enough to the MW data -- improved simulations and more accurate data are needed to confirm this interpretation. 
Thereby, no robust statistical conclusions can be drawn by the moment.


\subsection{Comparison to other works}
\label{ch4:sec:comparison_works}

Our results on timescales compare satisfactorily well with results from the analyses of the two zoom-in simulations presented in Paper IV, see Fig.~\ref{ch4:fig:CDF-timescales}.
This agreement adds robustness to our numerical results, as the 3 cosmological simulations involved are different in most respects, e.g. initial conditions, the integration methods used, and the subresolution modelizations implemented.

This work presents the first study of a statistical sample of long-lasting kinematically coherent-selected planes of satellites around MW/M31-like hosts.
Concerning comparisons with results of other cosmological simulation works,
it is worth noting that our unique methodology precludes exact comparisons.
For example, we recall that we study long-lasting planes, by requiring satellite orbital pole quasi-conservation (i.e., a variation within an aperture of $\alpha_{\rm co-orbit}$) along long time intervals of cosmic evolution,
while many studies just ask kinematic coherence at $z=0$.
Moreover, our 24\% frequency of MW/M31-like hosts with KPPs
are based on our specific KPP and satellite selection criteria, and hence no strict comparisons can be made with other authors' results.

A tendency for finding higher fractions of HS systems hosting kinematic or positional satellite planes seems to emerge in the very recent literature.
Several of these studies are devoted to analyze the TNG50 simulation at $z=0$.
\citet{Xu23} investigated MW-like hosts with over 14 satellites, finding just one  analog to the MW satellite plane in both spatial and kinematic properties  (excluded from the P24 sample) with 11 satellites showing conserved orbital poles over time.
\citet{HuTang24} extended these analyses.
They found that a $\sim 11.3\%$ of systems  hosted a plane of satellites, with this fraction increasing up to $\sim30\%$ for MW analogs at $z=0$. 
However, no analysis of their persistence in time was carried out.
\citet{Seo2024} studied the so-called `rareness' of the MW's plane of satellites,  finding it to  remain rare ($\sim 2.48\%$) in terms of the $c/a$ ratio distribution calculated from the 11 brightest satellites orbiting  around  202 MW-like galaxies. 
This is consistent with our results concerning the MW plane of satellites to the extent that their comparison makes sense.

\citet{Uzeirbegovic2024} develop the so-called 'planarity' tool.
They apply this method to the \textit{GaiaEDR3} data of MW satellites, and to 20 MW analogues from the \textsc{NewHorizon} cosmological simulations, where they found that 90\% of these systems show high planarity values both in positional and velocity spaces, while planarity is not supported for MW satellites in velocity space. 
See, however, \citet{Pawlowski2024planarity}.

\citet{Shao19} had already examined co-orbiting planes of satellites in MW-mass hosts within the EAGLE100 hydro-simulation ($m_{\rm DM}\sim10^6$M$_{\odot}$, 100 Mpc$^3$) \citep{schaye15}, identifying systems where at least 8 of the 11 most massive satellites orbit in a narrow plane at $z=0$, and they  reported a 13\% frequency of co-orbiting satellite systems.

A further note concerns the resolution of the simulations referenced above. In EAGLE100, the dark matter particle mass was $\sim20$ times higher than in TNG50. Apart from our inferences here, \citet{Sawala22} and \citet{Pham22} have shown that the frequencies of satellite planes are highly influenced by processes such as artificial disruption from numerical effects and disruption by the central host galaxy disk. Correcting for these issues (or with higher resolution) leads to higher probabilities to find MW-like satellite planes in $\Lambda$CDM, than compared to previous lower-resolution estimates.
Thereby, the sample frequency ($f_{\rm KPP-HS}^{\rm P24}$ in this work) of finding planes of satellites may only rise as we advance towards the next generation of large-volume, higher-resolution cosmological boxes.

Relative to these analyses, our work leads to similar qualitative results, within a different approach involving the persistence of kinematic planes and a careful statistical analysis of most of their important properties along cosmic evolution.


\section{Summary and conclusions}\label{ch4:sec:Summary}



In this paper we present the results obtained in our quest for Kinematically-Persistent Planes (KPPs) of satellites by applying the $\vec{J}_{\rm stack}$ method to the \citet{Pillepich24_MWM31} MW/M31-like sample of galaxies
(190 host-satellite `HS' systems, the so-called P24 sample, a subset of central galaxies of the TNG50 simulation).
TNG50 is  one of the highest resolution large-volume simulations to date.
The $\vec{J}_{\rm stack}$ method  was conducted from a  Universe age of T$_{\rm uni}\simeq6$ Gyr onward.
The method returns the direction of maximum satellite  co-orbitation, $\vec{J}_{\rm stack}$,  based on an orbital pole persistence condition along the said period.  For a given HS system, its KPP satellites  (temporarily co-orbiting satellites)  are then identified as those satisfying a persistent (temporal) co-orbitation condition with respect to $\vec{J}_{\rm stack}$. 
Within the selection criteria used in this work, we obtain a total of 46 KPP-HS systems (i.e. HS systems where a KPP has been identified).
%
%
%
Combined sets of KPP satellites and temporarily co-orbiting satellite sets have been studied under the name of ``co-orbiting satellite''  (CS) sets.
These results have enabled a statistical study of the frequency of kinematically persistent planes of satellites around MW–like galaxies in a $\Lambda$CDM context.

We have characterized the quality of orbital pole clustering of the detected KPPs (and CS sets) in terms of the so-called `pole co-orbitation excess' of their satellites. This is defined as the excess of orbital poles within a specific angular range of $\alpha_{\rm co-orbit}=36.^{\circ}87$ relative to a randomized distribution
(see appendix C in Paper III), normalized to its corresponding dispersion $\sigma$, with $\sigma$ being an $N_{\rm sat}$-dependent quantity.
We have computed this parameter for the CS sets 
($f_{\rm sat}^{\rm exc}(t)$), and for  the KPP satellites in the system ($f_{\rm KPP}^{\rm exc}$). 
Additionally, for each KPP-HS system, we  further analyzed the fraction of satellites co-rotating in one specific sense with respect to the total number of KPP satellites.

For each satellite, an orbital analysis has been carried out, allowing us to characterize its first apocenter (a kind of turn-around for satellites) and infall timescales, $T_{\rm apo1}^i$ and $T_{\rm inf}^i$, distinguishing between KPP and non-KPP satellite populations. We define the median values for each group as the $T_{\rm apo1}$ and $T_{\rm inf}$ timescales.

The evolution of the angular momentum of satellites relative to the host c.o.m., has been determined  from T$_{\rm uni}$ = 2 Gyr onwards. It turns out that KPP satellite orbital poles are conserved from very early times, while this is not always the case for satellites outside these structures. However, for KPP satellites, a persistent alignment with its respective  $\vec{J}_{\rm stack}$ co-orbitation axis   occurs only from  a Universe age termed  $T_{\rm cluster, i}^{\rm Jstack}$, for the $i$-th satellite in the set.
A timescale for the onset of orbital pole clustering, $T_{\rm cluster}^{\rm Jstack}$, has been defined as the median value of the individual timescales $T_{\rm cluster, i}^{\rm Jstack}$ for the satellites that are  KPP members (see also Paper IV).

KPPs (and CS sets) have also been characterized in terms of the three-dimensional structure defined by the positions of
their respective  satellite members. This has been done through a TOI analysis \citep{Cramer}.
We find that these planar configurations are generally thin and oblate, and centered at the host center, with only occasional deviations.
A relevant result is that the normal vector to the positional planar configuration of KPP satellite members, $\vec{n}_{\rm KPP}$, and the $\vec{J}_{\rm stack}$, remain parallel across time.
Exceptions occur for some KPP-HS systems during narrow time intervals, when $b(t) \sim c(t)$.
The same is true for the extended CS sets.
%

We have done a statistical analysis of the joint distribution of the $b/a$ and $c/a$ axis ratios, for positional planes formed both by KPP satellites and by CS sets. The triaxiality parameter $T$ has also been studied, concluding that the null-hypothesis of these statistical sets being indistinguishable cannot be rejected. A similar analysis has been done for the alignment between the normals $\vec{n}_{\rm KPP}$ and $\vec{n}_{\rm CS}$ obtaining that only 0.5\% of thin and oblate planes are misaligned.

Finally, the behavior of KPP satellite members in the velocity space has been analyzed through the kinematic morphological $\kappa_{\rm rot}(t)$ curve.


These are the main results and conclusions from this study on the frequency and properties of early KPPs (and CS sets) in the P24 sample:

\begin{enumerate}[label = {\arabic*)}, wide, left=0pt,labelindent=0pt]
\item The number of HS systems whose satellite population, $N_{\rm sat}$,  is high enough that the search for KPPs can be undertaken, given the selection criteria we impose to  both satellite galaxies and KPPs,  is $N_{\rm HS}^{\rm eli} = 123$. This search  gives a total of $N_{\rm HS}^{\rm KPP} = 46$  HS systems in the P24 sample hosting a KPP.

\item We quantify, for the first time, the frequency of  persistent kinematic planes around MW/M31-like hosts in $\Lambda$CDM.
The values for $N_{\rm HS}^{\rm eli}$ and  $N_{\rm HS}^{\rm KPP}$ shown above result  in a  total fraction $f_{\rm KPP-HS}^{\rm P24} \sim$ 24\% of HS systems hosting a KPP relative to the total sample of HS systems, 
and a fraction of $f_{\rm KPP-HS}^{\rm P24-eli} \sim  40\%$ relative to the eligible systems, i.e., where  the KPP search can be undertaken.

\item While the values of the $f_{\rm KPP-HS}^{\rm P24-eli}$ fraction do not seem to depend significantly  on the choice of the minimum satellite stellar mass ($M_{\rm *, cut}$, see Tab.~\ref{ch4:tab:FrequenciesTable}), it turns out that the $f_{\rm KPP-HS}^{\rm P24}$ fraction does depend on $M_{\rm *, cut}$. In particular, Tab.~\ref{ch4:tab:FrequenciesTable} indicates that this fraction increases when $M_{\rm *, cut}$  is halved (implying an increase of the average value of $N_{\rm sat}$ per host).
A possible implication is that $f_{\rm KPP-HS}^{\rm P24}$ {would rise} with increasing mass resolution in simulations,
as long as the selection criteria for both  hosts and KPPs are not changed.

\item HS systems  where no early KPPs have been identified (nonKPP-HS systems) exhibit delayed halo mass assembly histories and satellite infall timescales.
Their first, violent mass assembly phase lasts for longer than for KPP-HS systems, and, contrary to the latter, they generally suffer from important dynamical activity after T$_{\rm uni} \simeq 6$ Gyr, including mergers.
While KPP-HS systems form preferentially in populated/dense environments, nonKPP-HS systems tend to appear in less dense ones.

\item Satellite galaxies  belonging to KPPs are statistically distinguishable from satellites outside KPPs in a given KPP-HS systems. The former are, on average, further away from the host center at pericenter  and  have higher specific angular momenta than the latter. However, satellite subhalo turn-around or halo infall timescales, as well as  \textit{satellite infall mass} are not parameters determining satellite membership or not to a KPP.

\item Measuring their `pole clustering excess' parameter, it has been found that KPP satellites (CS sets) show a mean degree of orbital pole clustering around $\sim2$ (3)$\sigma$ in excess of that expected from a randomized distribution of poles.

\item Orbital pole clustering (and hence KPP formation) occurs during the fast phase of mass assembly of the host halo, i.e. prior to $T_{\rm no-fast}$, as well as before satellite infall time $T_{\rm inf}$. This implies that halo processes are not relevant in driving the clustering of KPP satellite orbital poles.
This result supports a scenario where early KPPs have a cosmological origin.

\item Satellite positions in KPPs and CS sets show oblate and planar configurations since T$_{\rm uni}\simeq6$ Gyr  over long periods of time. 
Within these periods of time, KPPs and CS planes exhibit strong alignment between their respective normal vectors $\vec{n}_{\rm KPP}$ and $\vec{n}_{\rm CS}$.
Moreover, both normal vectors align with $\vec{J}_{\rm stack}$.

\item KPP positional structures play the role of a kind of skeletons for positional planes formed by satellites in the extended CS sets. Indeed, the former shape, and determine the normal directions, of  the latter.

\item KPP satellites behave as disky structures according to the kinematic morphological parameter $\kappa_{\rm rot}$, adapted from \citet{Sales:2012}. Once satellites follow a common orbital plane motion, their orbits become more and more circular, eventually reaching a maximum stable value $\kappa_{\rm rot}^{\rm eq}$.

\item The degree of pole clustering excess over randomized pole distributions that we measure for MW satellites  is broadly consistent with the distribution of  values found for CS sets in TNG50. This degree of clustering is expressed in units of the respective dispersion of randomized poles, which somewhat corrects for the dependence in $N_{\rm sat}$.
Moreover,  the morphological parameter distribution of TNG50 planes includes the current MW values calculated with 21 kinematic satellites. However, given the paucity of P24 HS systems with as many satellites as considered for the MW, no robust statistical conclusions can be drawn at the moment.

\end{enumerate}

The analyses presented in this paper represent a significant advancement towards the quantification of the frequency, in a cosmological context, of MW/M31-like systems hosting KPPs.
Expanding upon the work conducted in Papers III and IV, which focused on two zoom-in simulations, this study extends the investigation to larger cosmological volumes with higher numerical and spatial resolutions.
Our findings indicate that kinematically-coherent planes of satellites are more frequent than previously reported, alleviating the tension between $\Lambda$CDM and MW satellite plane observations.
Moreover, we statistically proved that KPPs act as a kind of skeleton, shaping positional planes of CS sets and determining their normal directions.
This is particularly noteworthy given that, unlike the isolated environments examined in those earlier studies, the extension to a large cosmological box encompasses a wider sample of environments surrounding MW/M31-like systems.

Identification of KPP structures in a large-volume simulation allows for a more precise scenario of the conditions conducive to the formation and destruction of KPPs.
In a forthcoming paper (Gámez-Marín et al. in prep.), we will further explore, and delve deeper into, the origin of early KPPs, focusing on the impact of local environment factors shaping these systems. This forthcoming study aims to deepen our understanding of the early formation processes of satellite systems as well as the broader implications for their evolution in various galactic contexts.

\section*{Acknowledgements}
We thank the Ministerio de Ciencia e Innovación (Spain)  for financial support under Project grant PID2021-122603NB-C21.
M.G.M. acknowledges support from the MINECO/FEDER funding (Spain) through a FPI fellowship associated to PGC2018-094975-C21 grant, and thanks Dr. Pedro Cataldi for his help and comments with the merger trees.
I.S.S. acknowledges support by the European Research
Council (ERC) through Advanced Investigator grant to C.S. Frenk, DMIDAS (GA 786910), and from the Science and Technology Facilities Council STFC ST/P000541/1 and ST/X001075/1.
S.E.P. acknowledges support from MinCyT (Argentina) through BID PICT 202000582.
This work has received financial support from the European Union's HORIZON-MSCA-2021-SE-01 Research and Innovation programme under the Marie Sklodowska-Curie grant agreement number 101086388 - Project acronym: LACEGAL.
The authors are grateful to Dr. D. Sotillo-Ramos for his valuable advice regarding the use of the TNG50 database, and to Dr. A. Pillepich, Dr. D. Nelson, and their team for making publicly available the data from their TNG50 sample of MW/M31-like galaxies.

\vspace{-0.5cm}
\section*{Data Availability}
The simulation data used in this article is publicly available and accessible at \url{https://www.tng-project.org/data/downloads/TNG50-1/}. The analysis code underlying this work may be shared upon request within a collaboration framework with the authors.

\bibliography{bib_THESIS_MGM}{}
\bibliographystyle{mnras}

\appendix
\section{Satellite radial distance evolution}
\label{ch4:appendixA}

\begin{figure}
\centering
\includegraphics[width=\linewidth]{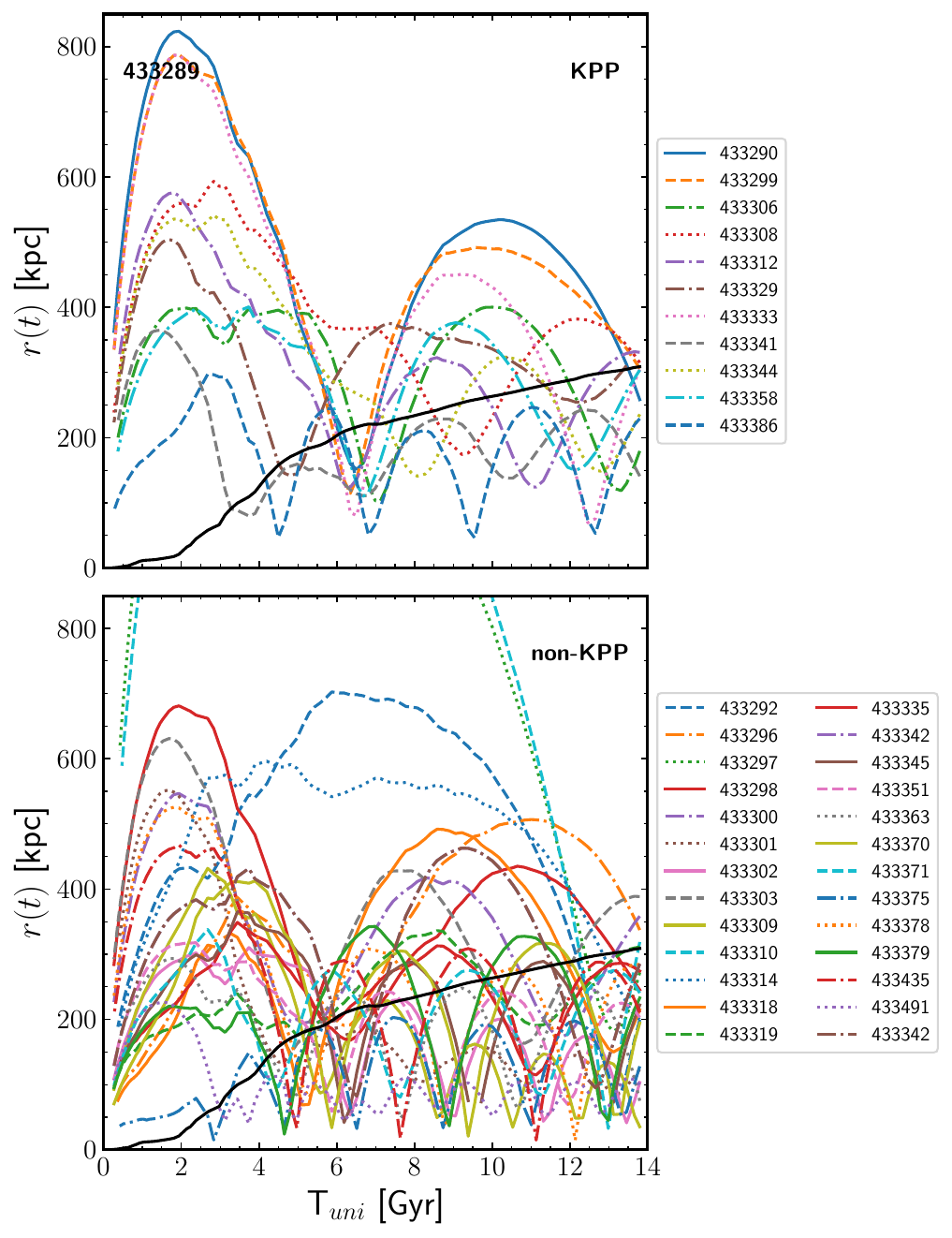}
\caption{
Evolution of the radial distance with respect to the host center, $r(t)$,  for satellites belonging to the \#433289 KPP-HS system (see Tab.~\ref{ch4:tab:DataTable1}).
Codes for line types and colours are given in legends on the right of each panel.
\textit{Upper (lower) panel}: KPP (non-KPP) satellite members.
Black continuous lines: host virial radius, $R_{\rm 200c}(t)$.
}
\label{ch4:fig:RadDistWithSubSat-Evol}
\end{figure}

Fig.~\ref{ch4:fig:RadDistWithSubSat-Evol} shows the evolution of the radial distances of satellites to their host c.o.m. in one example KPP-HS system. We split the satellite population into KPP satellite members (upper panel), and non-KPP satellites (lower panel). 
The black, continous lines stand for the time evolution of the host halo's virial radius, $R_{\rm 200c}(t)$. Note that each satellite maintains consistent color and line codes.
Some common characteristics of the $r(t)$ curves are clear from the two panels of Fig.~\ref{ch4:fig:RadDistWithSubSat-Evol}: satellites first show $r(t)$ increasing curves, until they  reach a maximum (the first apocenter distance at $T_{\rm apo1}^i$).  Then the $r(t)$ curve  decreases, crosses  the $R_{\rm 200c}$ at T$_{\rm uni}=T_{\rm inf}^i$, where  it reaches the first pericentric distance and begins to orbit around its host. Most satellites show from this time onwards roughly constant periods and apocentric distances.
KPP satellites show statistically-significant larger pericenters than non-KPP satellites (see Fig.~\ref{ch4:fig:DpNorm-Dist}). 

\vspace{-0.4cm}

\section{Dependence of morphological planes of CS sets on the number of satellites}
\label{ch4:appendixB}

As explained in Secs.~\ref{ch4:sec:Quant_SatCoorb} and \ref{ch4:sec:KPP_PosSpace} of the main text (see also Sec.~\ref{ch4:sec:Comparison_MW}) to  make  statistical analyses on measures  of satellite co-orbitation and on positional planes properties,  we  use the information contained in the 2254 CS planes of the so-called \textit{statistical sample of CS sets}.  Comparing the results of these analyses to MW data \citep[50 satellites, with 21 co-orbiting, see][]{Taibi24} would demand a statistical sample of CS sets whose total number of satellites ($N_{\rm sat}$) and co-orbiting number of satellites ($N_{\rm CS}$) are similar to those of the MW sample we compare with.

Only a few CS sets in the statistical sample reach values as high as $N_{\rm CS}=21$ and/or $N_{\rm sat}=50$, or slightly more, precluding in principle meaningful comparisons. 
A way out of this situation consists in proving that the values of satellite co-orbitation and positional plane properties for the statistical sample depend neither on $N_{\rm CS}$ and $N_{\rm sat}$. 
In such case, comparisons can be made with the statistical sample of CS sets, 
In such case, comparisons  can be made with the statistical sample of CS sets, and the conclusions can then be considered valid for higher $N_{\rm CS}$ and $N_{\rm sat}$ values.
To study these dependences, Spearman correlation coefficients have been calculated for the values of the co-orbitation measures and positional planes properties studied in this Paper, versus $N_{\rm sat}$ and/or $N_{\rm CS}$. Results are given in Tab.~\ref{ch4:tab:Correlations_Ncs} (for the pole clustering excess $f_{\rm sat}^{\rm exc}(t)$ vs $N_{\rm sat}$ plot we obtain $\rho=0.184$). They indicate that the dependence is at most weak when low values for $N_{\rm CS}$ are removed ($N_{\rm CS}\geq10$ and $N_{\rm CS}\geq15$). Thereby direct comparisons can be made. 
However, due to the scarcity of  CS sets with a number of satellites comparable to the MW data considered here, improved simulations and more precise observations are still needed to draw robust statistical conclusions from these comparisons.

\begin{table}
\caption{Spearman correlation coefficients measuring the dependence of the different morphological parameters of CS sets on the number of co-orbiting satellites, $N_{\rm CS}$, in the analyzed KPP-HS systems. Each row shows the correlation values when only considering CS sets with different $N_{\rm CS}$ minimum values.}
\scriptsize
\hspace{0cm}
\vspace{0cm}
\begin{tabular}{|l | c  | c | c | c | c |}
\toprule
\multicolumn{1}{|c|}{$N_{\rm CS}$} & \multicolumn{1}{|c|}{Total CS planes} & \multicolumn{1}{c|}{$T_{\rm CS}$} & \multicolumn{1}{c|}{$(b/a)_{\rm CS}$} & \multicolumn{1}{c|}{$(c/a)_{\rm CS}$} & \multicolumn{1}{c|}{$(f_{\rm co-rot})_{\rm CS}$}  \\
\hline
\multicolumn{1}{|c|}{$>0$} &  \multicolumn{1}{c|}{2254} & \multicolumn{1}{c|}{-0.312} & \multicolumn{1}{c|}{0.363} & \multicolumn{1}{c|}{0.491} & \multicolumn{1}{c|}{-0.031}  \\
\hline
\multicolumn{1}{|c|}{$\geq10$} & \multicolumn{1}{c|}{941}   & \multicolumn{1}{c|}{0.033} & \multicolumn{1}{c|}{-0.030} & \multicolumn{1}{c|}{0.120} & \multicolumn{1}{c|}{0.043}  \\
\hline
\multicolumn{1}{|c|}{$\geq15$} & \multicolumn{1}{c|}{330}   & \multicolumn{1}{c|}{0.015} & \multicolumn{1}{c|}{-0.014} & \multicolumn{1}{c|}{-0.014} & \multicolumn{1}{c|}{-0.231} \\
\bottomrule
\end{tabular}
\label{ch4:tab:Correlations_Ncs}
\end{table}

\clearpage
\section*{Glossary}

\noindent
\begin{tabularx}{\textwidth}{@{} l X @{}}
  \toprule
  \textbf{Term} & \textbf{Definition} \\
  \midrule
  P24 sample & Sample of 190 MW/M31-like central galaxies presented in \cite{Pillepich24_MWM31}. \\
  HS system &  Host-satellite system. \\
  $N_{\rm sat}$ &  Total number of satellites in an HS system. \\
  KPP & Kinematically-Persistent Plane -- Fixed sets of satellites whose orbital poles are conserved over long time intervals and whose orbital poles remain clustered, giving rise persistent-in-time planes. \\
  KPP-HS systems & Host-satellite systems hosting a KPP. \\
  nonKPP-HS systems & Host-satellite systems not hosting a KPP, according to our KPP selection criteria. \\
  $N_{\rm KPP}$ & Number of satellites forming the KPP in a KPP-HS system. \\
  KPP satellites & Satellites forming the KPP in a given KPP-HS system. \\
  non-KPP satellites & Satellites that are not part of the KPP in a given KPP-HS system. \\
  $f_{\rm sat}$ & Fraction of satellites with respect to the total number of satellites in a given HS system. \\
  $f_{\rm KPP}$ & Fraction of satellites forming the KPP in a KPP-HS system. \\
  $f_{\rm KPP-HS}^{\rm P24}$ & Fraction of HS systems where a KPP has been identified over the total number of HS systems in the P24 sample. \\
  $f_{\rm KPP-HS}^{\rm P24-eli}$ & Fraction of HS systems where a KPP has been identified over the total number of \textit{eligible} HS systems in the P24 sample. \\
  $f_{\rm co-rot}$ & Fraction of satellites co-rotating in one sense with respect to the total number of KPP satellites ($N_{\rm KPP}$), see Sec.~\ref{ch4:sec:Quant_SatCoorb}. \\
  $\vec{J}_{\rm stack}$ & Direction of maximum satellite co-orbitation. This is the axis that encloses the maximum number of satellite orbital poles within an angular aperture $\alpha_{\rm co-orbit}=36.^{\circ}87$. It defines the orbital plane of the KPP. \\
  $f_{\rm sat}^{\alpha_{\rm co-orbit}}$ & Fraction of satellites with orbital poles within an angular distance $\alpha\leq\alpha_{\rm co-orbit} = 36.^{\circ}87$ from $\vec{J}_{\rm stack}$. \\
  $f_{\rm rand}^{\alpha_{\rm co-orbit}}$ & Mean fraction of satellites with a randomized distribution of orbital poles at an angular distance $\alpha\leq\alpha_{\rm co-orbit} = 36.^{\circ}87$ from $\vec{J}_{\rm stack}$. \\ 
  $\sigma$ & Dispersion value of the fraction of satellites whose randomized distribution orbital poles are at an angular distance from $\vec{J}_{\rm stack}$ $\alpha\leq\alpha_{\rm co-orbit} = 36.^{\circ}87$. \\
  $f_{\rm KPP}^{\rm exc}$ & Orbital pole co-orbitation excess for the fraction $f_{\rm KPP}$ of KPP satellites with respect to a randomized distribution of orbital poles. See Sec.~\ref{ch4:sec:Quant_SatCoorb} for details. \\
  $f_{\rm sat}^{\rm exc}$ & Orbital pole co-orbitation excess for the fraction $f_{\rm sat}^{\alpha_{\rm co-orbit}}(t)$ of co-orbiting satellites at T$_{\rm uni}=t$ with respect to a randomized distribution of orbital poles. See Sec.~\ref{ch4:sec:Quant_SatCoorb} for details. \\
  CS set & Co-orbiting satellites -- satellites that have their orbital poles aligned with the $\vec{J}_{\rm stack}$ axial vector, either temporarily or persistently. \\
  $\vec{n}_{\rm KPP}$ & Normal vector to the positional plane adjusted to the KPP satellites in a given KPP-HS system. \\
  $\vec{n}_{\rm CS}$ & Normal vector to the positional plane adjusted to the CS sets in a given KPP-HS system at a specific T$_{\rm uni}=t$. \\
  $\kappa_{\rm rot}^{\rm eq}$ & Maximum stable value of the kinematic morphological parameter $\kappa_{\rm rot}(t)$ for a given KPP-HS system. \\
  $M_{\rm subh}$ & Total mass of all member particle/cells bound to a subhalo, of all types. \\
  \bottomrule
\end{tabularx}


\bsp	
\label{lastpage}
\end{document}